\documentclass[nofootinbib,aps,11pt]{revtex4-1}
\usepackage{graphicx}
\usepackage{hyperref}
\usepackage{cancel}
\usepackage{amssymb}
\usepackage{textcomp}
\usepackage{amsmath}
\usepackage{bm}
\usepackage{times}
\usepackage{epsfig}
\usepackage{color}
\usepackage[utf8]{inputenc}
\usepackage{mathrsfs}
\usepackage{esint}
\usepackage{ulem}
\usepackage{subfigure}
\usepackage{makecell}
\usepackage{booktabs}
\usepackage{multirow}

\newcommand{\GeV}{{\rm GeV}}
\newcommand{\TeV}{{\rm TeV}}

\begin{document}
\title{\LARGE Predictive Scotogenic Model with Flavor Dependent Symmetry}
\bigskip
\author{Zhi-Long Han$^1$}
\email{sps\_hanzl@ujn.edu.cn}
\author{Weijian Wang$^2$}
\email{wjnwang96@aliyun.com}
\affiliation{
$^1$School of Physics and Technology, University of Jinan, Jinan, Shandong 250022, China
\\
$^2$Department of Physics, North China Electric Power University, Baoding 071003, China
}
\date{\today}

\begin{abstract}
 In this paper, we propose a viable approach to realise two texture-zeros in the scotogenic model with flavor dependent $U(1)_{B-2L_\alpha-L_\beta}$ gauge symmetry. These models are extended by two right-handed singlets $N_{Ri}$ and two inert scalar doublets $\eta_{i}$, which are odd under the dark $Z_2$ symmetry. Among all the six constructed textures, texture $A_1$ and $A_2$ are the only two allowed by current experimental limits. Then choosing texture $A_1$ derived from $U(1)_{B-2L_e-L_\tau}$, we perform a detail analysis on the corresponding phenomenology such as predictions of neutrino mixing parameters, lepton flavor violation, dark matter and collider signatures. One distinct nature of such model is that the structure of Yukawa coupling $\bar{L}\tilde{\eta}N_R$ is fixed by neutrino oscillation data, and can be further tested by measuring the branching ratios of charged scalars $\eta_{1,2}^\pm$.
\end{abstract}

\maketitle

\section{Introduction}
It is well known that the Standard Model (SM) needs extensions to accommodate two missing spices: the tiny but no-zero neutrino masses and the cosmological dark matter (DM) candidates. One way of incorporating above two issues in a unified framework is the scotogenic model \cite{Ma:2006km,Krauss:2002px,Aoki:2008av}, where neutrinos are radiatively generated and the DM fields serves as intermediate messengers propagating inside the loop diagram. With all new particles around TeV scale, the scotogenic model leads to testable phenomenologies \cite{Ma:2006fn,Hambye:2006zn,Sierra:2008wj,Suematsu:2009ww,Schmidt:2012yg,
Bouchand:2012dx,Ma:2012ez,Klasen:2013jpa,Ho:2013hia,Modak:2014vva,
Molinaro:2014lfa,Faisel:2014gda,Merle:2015gea,Ahriche:2016cio,Lindner:2016kqk,
Hessler:2016kwm,Borah:2017dfn,Abada:2018zra,Hugle:2018qbw,Baumholzer:2018sfb,
Borah:2018rca,Bian:2018bxr}. Therefore, viable models are extensively studies in recent years \cite{Cai:2017jrq}.

On the other hand, the understanding of the leptonic flavor structure is still one of the major open questions in particle physics. The consensus is that the leptonic mass texture is tightly restricted under the present experimental data. An attractive approach is to consider two texture-zeros in neutrino mass matrix ($M_{\nu}$) so that the number of parameters in the Lagrangian is reduced\cite{Frampton:2002yf}. The phenomenological analysis of two texture-zeros models have been studied in Ref.\cite{Fritzsch:2011qv,Alcaide:2018vni}.  Among fifteen logically patterns, seven of them are compatible to the low-energy experimental data.

On the theoretical side, the simplest way of realizing texture-zeros is to impose the discrete $Z_{N}$ flavor symmetry\cite{Grimus:2004hf}. However,
it might be more appealing to adopt gauge symmetries instead of discrete ones,  because the latter may be treated as the residual of $U(1)$ gauge symmetry. It is noted that one can not set any restriction on lepton mass matrix by means of fields with flavor universal charges. Thus the flavor dependent $U(1)$ gauge symmetry is the reasonable choice. Along this thought of idea, specific models are considered in the context of seesaw mechanisms. In Ref.\cite{Araki:2012ip}, the two texture-zeros are realized based on the anomaly-free $U(1)_{X}$ gauge symmetry with $X\equiv B-\sum x_{\alpha}L_{\alpha}(\alpha=e,\mu,\tau)$ being the linear combination of baryon number $B$ and the lepton numbers $L_{\alpha}$ per family. In Ref.\cite{Cebola:2013hta}, more solutions are found in the type-I  and/or III seesaw framework.

It is then natural to ask if predictive texture-zeros in $M_{\nu}$ can be realized in the scotogenic scenario and several attempts have been made in this direction. For example, one texture-zero is recently considered in Ref.~\cite{Kitabayashi:2018bye}. Texture $B_1$-$B_4$ have been discussed in a model-independent way in Ref.~\cite{Kitabayashi:2017sjz}. Texture $C$ is obtained by introducing $U(1)_{L_{\mu}-L_{\tau}}$ gauge symmetry \cite{Baek:2015mna,Baek:2015fea,Lee:2017ekw,Asai:2018ocx}.  Texture $B_2$ is realised with $U(1)_{L_e+L_\mu-L_\tau}$ gauge symmetry in Ref.~\cite{Nomura:2017ohi}. If the quark flavor is also flavor dependent, e.g., $U(1)_{xB_3-xL_e-L_\mu+L_\tau}$, then one can further interpret the $R_{K}$ anomaly with texture $A_1$ \cite{Ko:2017quv}. Other viable two texture-zeros are systematically realised in Ref.~\cite{Nomura:2018rvy} by considering the $U(1)_{B-2L_\alpha-L_\beta}$ gauge symmetry with three right-handed singlets. In this paper, we provide another viable approach. Under same flavor dependent $U(1)_{B-2L_\alpha-L_\beta}$ gauge symmetry, we introduce only two right-handed singlets but two inert scalars, leading to different texture-zeros. In aspect of predicted phenomenology, the texture $B_1$ considered in Ref.~\cite{Nomura:2018rvy} is marginally allowed by current Planck result for $\sum m_i<0.12$ eV \cite{Aghanim:2018eyx}, we thus consider texture $A_1$ with latest neutrino oscillation data \cite{Esteban:2018azc} as the benchmark model. In this case, the gauge symmetry is $U(1)_{B-2L_e-L_\tau}$ in our approach.

The rest of this paper is organised as follows. Start with classic scotogenic model in Sec.~\ref{Sec:MD}, we first discuss the realization of texture-zeros in scotogenic model with $U(1)_{B-2L_\alpha-L_\beta}$ gauge symmetry in a general approach. Then the texture $A_1$ derived from $U(1)_{B-2L_e-L_\tau}$ is explained in detail. The corresponding phenomenological predictions, such as neutrino mixing parameters, lepton flavor violation rate, dark matter and highlights of collider signatures are presented in Sec.~\ref{Sec:PH}. Finally, conclusions are summarised in Sec.~\ref{Sec:CL}.

\section{The Model Setup}\label{Sec:MD}

\begin{figure}
\begin{center}
\includegraphics[width=0.4\linewidth]{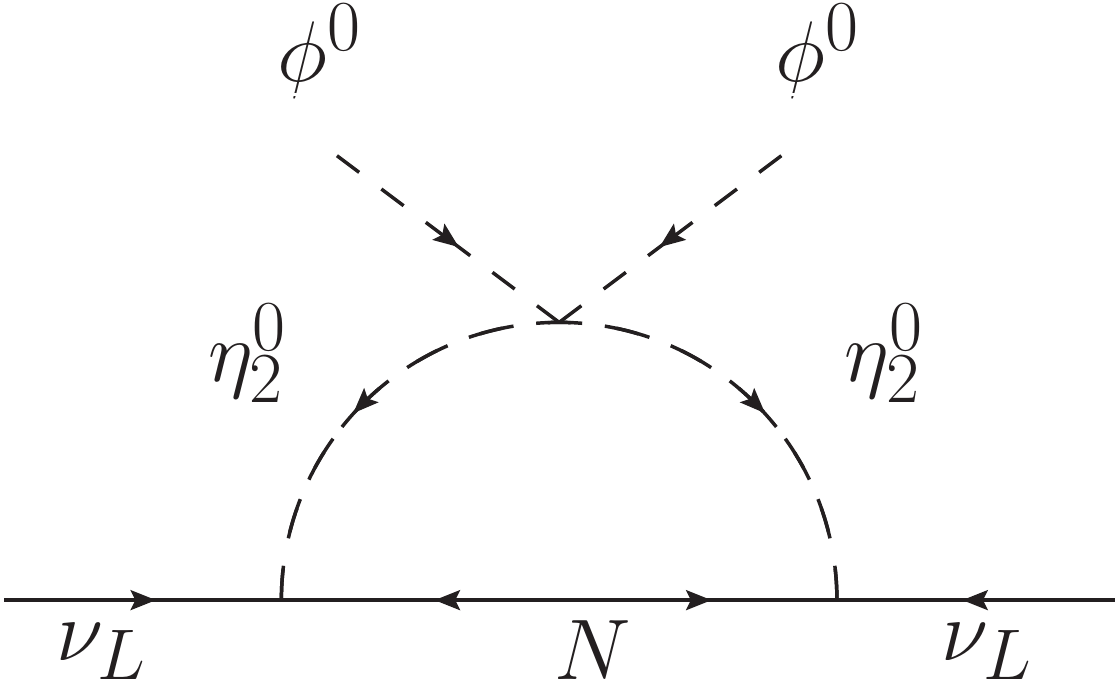}
\includegraphics[width=0.4\linewidth]{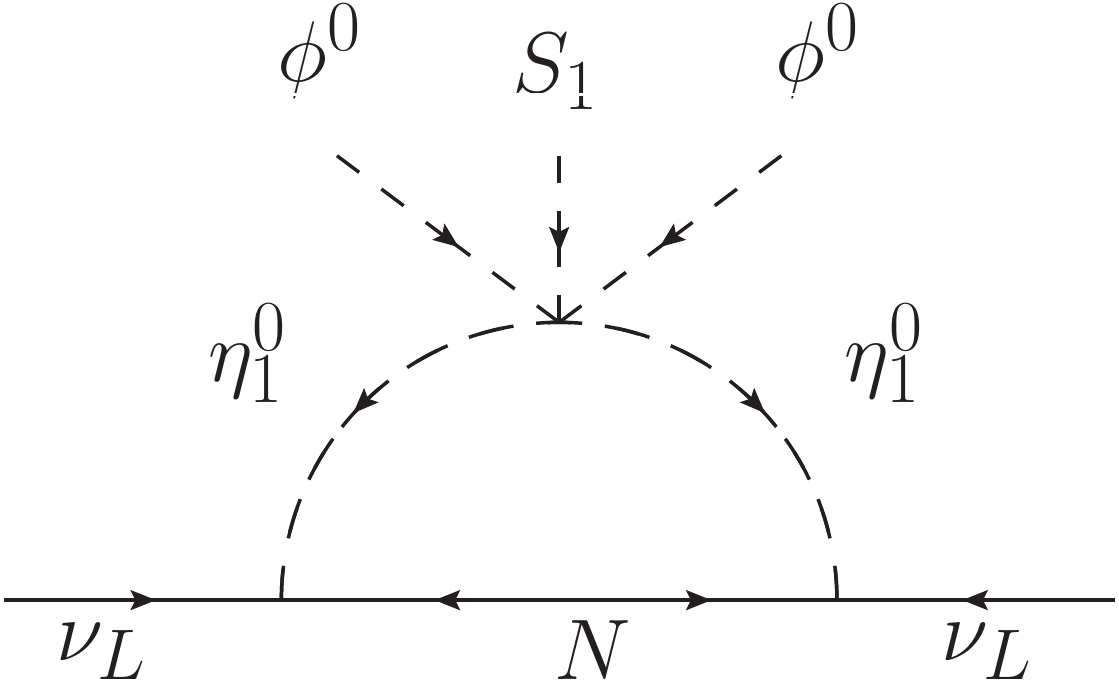}
\end{center}
\caption{Radiative neutrino mass at one-loop. Left pattern is for classic scotogenic model, while right pattern is the additional contribution in our models.
\label{Fig:FR}}
\end{figure}

\subsection{Classic scotogenic model}

In the classic scotogenic model proposed by Ma \cite{Ma:2006km}, three right-handed fermion singlets $N_{Ri}(i=1\sim3)$ and an inert scalar doublet field $\eta=(\eta^{+},\eta^{0})$ are added to the SM.  In addition, a discrete $Z_{2}$ symmetry is imposed for the new fields in order to forbid the tree-level neutrino Yukawa interaction and stabilize the DM candidate. The relevant interactions for neutrino masses generation are given by
\begin{equation}\label{ma}
\mathcal{L} ~\supset~ h_{\alpha i}\overline{L}_{\alpha}\tilde{\eta} N_{Ri}+\frac{1}{2}
 M_{N}\overline{N}_{R}^{c}N_{R}+\frac{1}{2}\lambda
(\Phi^{\dagger}\eta)^2+\text{h.c.}.
\end{equation}
The mass matrix $M_{N}$ can be diagonalized by an unitary matrix $V$ satisfying
\begin{equation}
V^{T}M_{N}V=\hat{M}_{N}\equiv\text{diag}(M_{N1},M_{N2},M_{N3}).
\end{equation}
Due to the $Z_{2}$ symmetry, the neutrino masses are generated at one-loop level, as show in left pattern of Fig.~\ref{Fig:FR}. The neutrino mass matrix can be computed exactly, i.e.
\begin{equation}
(M_{\nu})_{\alpha\beta}=\frac{1}{32\pi^{2}}\sum_{k}
h_{\alpha i}V_{ik}h_{\beta j}V_{jk}M_{Nk}\Big[\frac{m_{R}^{2}}{m_{R}^{2}\!-M_{Nk}^{2}}
\log\big(\frac{m_{R}^{2}}{M_{Nk}^{2}}\big)-\frac{m_{I}^{2}}{m_{I}^{2}\!-M_{Nk}^{2}}
\log\big(\frac{m_{I}^{2}}{M_{Nk}^{2}}\big)\Big]\\
\end{equation}
where $m_{R}$ and $m_{I}$ are the masses of $\sqrt{2}\Re \eta^0$ and  $\sqrt{2}\Im \eta^0$. If we assume $m_{0}^2\equiv(m_{R}^2+m_{I}^2)/2\gg M_{Nk}^2$, $M_{\nu}$ are then given by
\begin{equation}\label{scale}\begin{split}
(M_{\nu})_{\alpha\beta}&\simeq -\frac{1}{32\pi^{2}}\frac{\lambda v^2}{m_{0}^2}\sum_{k}h_{\alpha i}V_{ik}h_{\beta j}V_{jk}M_{Nk}\\
&=-\frac{1}{32\pi^{2}}\frac{\lambda v^2}{m_{0}^2}(hM_{N}h^{T})_{\alpha\beta}
\end{split}\end{equation}
The neutrino mass matrix $M_\nu$ is diagonalized as
\begin{equation}
U_\text{PMNS}^T M_\nu U_\text{PMNS}=\hat{m}_\nu\equiv\text{diag}(m_1,m_2,m_3),
\end{equation}
where $U_\text{PMNS}$ is the neutrino mixing matrix denoted as
\begin{align}
U_{\text{PMNS}}\! =\! \left(
\begin{array}{ccc}
c_{12} c_{13} & s_{12} c_{13} & s_{13} \\
-c_{12}s_{23}s_{13}-s_{12}c_{23}e^{-i\delta} & -s_{12}s_{23}s_{13}+c_{12}c_{23} e^{-i\delta} & s_{23}c_{13}\\
-c_{12}c_{23}s_{13}+s_{12}s_{23}e^{-i\delta} & -s_{12}c_{23}s_{13}-c_{12}s_{23}e^{-i\delta} & c_{23}c_{13}
\end{array}
\right)\!\times\!
\text{diag}(e^{i \rho},e^{i\sigma},1)
\end{align}
Here, we define $c_{ij}=\cos\theta_{ij}$ and $s_{ij}=\sin\theta_{ij}$ ($ij=12,23,13$) for short, $\delta$ is the Dirac phase and $\rho,\sigma$ are the two Majorana phases as in Ref.~\cite{Fritzsch:2011qv}.

\begin{center}
\begin{table}
\begin{tabular}{|c||c|c|c|c|c|c|c|c||c|c|c|c|c|}
\hline\hline
\multirow{2}{*}{Group} &\multicolumn{8}{c||}{Lepton Fields} & \multicolumn{5}{c|}{Scalar Fields} \\
 \cline{2-14}
 &~$L_\alpha$~ & ~$\ell_{\alpha R}$~& ~$L_\beta$ ~ &~$\ell_{\beta R}$ ~ &~$L_\gamma$ ~&~$\ell_{\gamma R}$ ~&~$N_{R1}$ ~&~$N_{R2}$~& ~$\Phi$ ~ & ~$\eta_{1}$~  & ~$\eta_{2}$ ~& ~$S_1$~  & ~$S_2$ \\ \hline
~$SU(2)_L$& ~$2$~ & ~$1$~& ~$2$ ~ &~$1$ ~ &~$2$ ~&~$1$ ~&~$0$ ~&~$0$~& ~$2$ ~ & ~$2$~  & ~$2$ ~& ~$1$~  & ~$1$
\\ \hline
~$U(1)_{Y}$& ~$-\frac{1}{2}$~ & ~$-1$~ & ~$-\frac{1}{2}$~ & ~$-1$~ & ~$-\frac{1}{2}$~ & ~$-1$~ & ~$1$~ & ~$1$~ & ~$\frac{1}{2}$~ & ~$\frac{1}{2}$~ & ~$\frac{1}{2}$~ & ~$0$~ & ~$0$~
\\ \hline
~$Z_2$& ~$+$~ & ~$+$~& ~$+$ ~ &~$+$ ~ &~$+$ ~&~$+$ ~&~$-$ ~&~$-$~& ~$+$ ~ & ~$-$~ & ~$-$ ~ & ~$+$~  & ~$+$
\\\hline
~$U(1)_{B-2L_\alpha-L_\beta}$& ~$-2$~ & ~$-2$~ &~$-1$ ~&~$-1$  & ~$0$ ~ &~$0$ ~&~$-1$ ~&~$-2$~& ~$0$ ~ & ~$-1$~ & ~$0$ ~ & ~$2$~  & ~$3$
\\\hline \hline
\end{tabular}
\caption{Particle content and corresponding charge assignments.}
\label{TB:Charge}
\end{table}
\end{center}

\subsection{Two texture-zeros in scotogenic model}

In this section, we demonstrate a class of scotogenic models with $G_{SM}\times U(1)_{B-2L_{\alpha}-L_{\beta}}\times Z_{2}$ gauge symmetry where two texture-zero structures in $M_{\nu}$ are successfully realized. The particle content and corresponding charge assignments are listed in Tab.~\ref{TB:Charge}. In the fermion sector, we introduce two right-handed $SU(2)_L$ singlets $N_{R1}$ and $N_{R2}$ and assume they carry the same no-zero $B-2L_{\alpha}-L_{\beta}$ charges as two of SM leptons respectively. Noticeably, if one further introduce one additional $N_{R3}$ with zero $B-2L_{\alpha}-L_{\beta}$ charge, the approach considered in Ref.~\cite{Nomura:2018rvy} are then reproduced.  In terms of gauged $U(1)_{B-2L_{\alpha}-L_{\beta}}$ symmetry, the anomaly free conditions should be considered first and we find all anomalies are zero because
\begin{eqnarray}
[SU(3)_{c}]^2U(1)_{X} &:& 3\times\frac{1}{2}\Big(\frac{2}{3}-\frac{1}{3}-\frac{1}{3}\Big)=0
\\ \nonumber
U(1)_{Y}[U(1)_{X}]^2 &:&3\Big[6\Big(\frac{1}{6}\Big)\!-3\Big(\frac{2}{3}\Big)\!-3\Big(\frac{1}{3}\Big) \Big]\Big(\frac{1}{3}\Big)^2\!\!+\Big[2\Big(\!-\frac{1}{2}\Big)\!-\!\Big(\!-1 \Big)\Big]\Big[(-1)^2+(-2)^2\Big]=0\\ \nonumber
[SU(2)_{L}]^2U(1)_{X} &:&  \frac{1}{2}\Big[3\times3\Big(\frac{1}{3}\Big)+(-1)+(-2)\Big]=0
\\ \nonumber
[U(1)_{Y}]^2U(1)_{X} &:&  3\Big[6\Big(\frac{1}{6}\Big)^2-3\Big(\frac{2}{3}\Big)^2-3\Big(\frac{1}{3}\Big)^2 \Big]\Big(\frac{1}{3}\Big)+\Big[2\Big(-\frac{1}{2}\Big)^2-\Big(-1 \Big)^2\Big](-1-2)=0
\\ \nonumber
U(1)_{X}^{3} &:& 2(-1)^3-2(-1)^3+2(-2)^3-2(-2)^3=0
\\ \nonumber
[\text{Gravity}]^2 U(1)_{X} &:&   2(-1)-2(-1)+2(-2)-2(-2)=0
\end{eqnarray}

Let us now discuss the scotogenic realizations of two texture-zeros in $M_{\nu}$. With two $N_{R}$ components,  $h$ and $M_{N}$ are $3\times 2$ and $2\times 2$ matrices respectively.  From Eq.\eqref{scale}, it is clear that the texture-zeros of $M_{\nu}$ can be attributed to the texture-zeros in $h$ and $M_{N}$ matrices. In the original scotogenic model with an inert scalar doublet $\eta$ and two $N_{R}$ fields, the charge assignments for  $U(1)_{B-2L_{\alpha}-L_{\beta}}$ gauge symmetry give rise to only two Yukawa terms for  $h_{\alpha i}\overline{L}_{\alpha}\tilde{\eta}N_{Ri}(\alpha=e,\mu,\tau,i=1,2)$. In this case, at least two texture-zeros are placed in the same line of $h$ matrix, being therefore excluded experimentally. In order to accommodate the realistic neutrino mixing data, the scotogenic model are extended where, in scalar sector, two inert doublet $\eta_{1}$ and $\eta_{2}$ are introduced (see Tab. \ref{TB:Charge}). In addition, two scalar singlet $S_{1}$ and $S_{2}$ are added so that $U(1)_{B-2L_{\alpha}-L_{\beta}}$ symmetry is spontaneously breaking after $S_{1,2}$ get the vacuum expectation value (VEV) $\langle S_{1,2}\rangle=v_{1,2}/\sqrt{2}$. Note that $N_{Ri}$ and $\eta_{i}$ are odd under the discrete $Z_{2}$ symmetry. Since we have two inert scalars, the relevant scalar interactions for the loop-induced neutrino masses is given by
\begin{equation}\label{es}
\mathcal{L}_S ~\supset \frac{\lambda}{\Lambda}(\Phi\eta_{1})^2S_{1}
+\lambda^{\prime}(\Phi\eta_2)^2+\text{h.c.},
\end{equation}
where $\Lambda$ is a new high energy scale and the first term is a dimension-five operator guaranteed by the accidental $U(1)_{B-2L_{\alpha}-L_{\beta}}$ symmetry. One can achieve the effective operator by simply adding a new scalar singlet $\rho\sim(1,0,1,-)$ so that in scalar sector $\mathcal{L}_S ~\supset\mu (\Phi^{\dagger}\eta)\rho^{\dagger}+\mu^{\prime}\rho^2 S_{1}$ is allowed. Then the effective interaction $\lambda(\Phi\eta_{1})^2S_{1}/\Lambda$ is obtained by integrate the $\rho$ field out of $\mathcal{L}_S$ sector. In the following analysis, we adopt the expression of effective operator in Eq.\eqref{es} and do not consider its specific realization in detail.

The neutrinos acquire their tiny masses radiatively though the one-loop diagram depicted in Fig.~\ref{Fig:FR}. Therefore, the neutrino mass matrix is formulated by two different contribution, namely,
\begin{equation}\label{mv}
(M_{\nu})\propto hM_{N}h^{T}+fM_{N}f^{T},
\end{equation}
where $h$ and $f$ are the Yukawa coupling texture for $\overline{L}_{\alpha}\tilde{\eta}_{1} N_{Ri}$ and $\overline{L}_{\alpha}\tilde{\eta}_{2} N_{Ri}$ with further assumption $\Lambda=\langle S_1 \rangle$ and $\lambda=\lambda'$. As a case study, we consider the $U(1)_{B-2L_{e}-L_{\tau}}$ gauge symmetry under which the flavor dependent Yukawa interaction is given by
\begin{eqnarray}\label{Eq:Yuk}
-\mathcal{L}_\text{Y} &=& h_{\mu1} \bar{L}_\mu \tilde{\eta}_1 N_{R1} + h_{\tau 2} \bar{L}_\tau \tilde{\eta}_1 N_{R2} + f_{\tau1} \bar{L}_\tau \tilde{\eta}_2 N_{R1}
+f_{e2} \bar{L}_e \tilde{\eta}_2 N_{R2} \\ \nonumber
&& + y_{11} \overline{N^c_{R1}} N_{R1} S_1 + y_{12} (\overline{N^c_{R1}} N_{R2}
 +\overline{N^c_{R2}} N_{R1}) S_2 + \text{h.c.},
\end{eqnarray}
where from the charge assignment, the texture of fermion Yukawa coupling are
\begin{align}\label{Eq:hfy}
h=\left(
\begin{array}{cc}
0 & 0\\
h_{\mu 1} & 0\\
0 & h_{\tau 2}
\end{array}\right),\quad
f=\left(
\begin{array}{cc}
0 & f_{e2}\\
0 & 0\\
f_{\tau 1} & 0
\end{array}\right),\quad
y=\left(
\begin{array}{cc}
y_{11} & y_{12}\\
y_{12} & 0
\end{array}\right).
\end{align}

\begin{center}
\begin{table}
\begin{tabular}{|c|c||c|c||c|}\hline\hline
 Texture of $M_{\nu}$ &Group &Texture of $M_{\nu}$ &Group& Status\\
\hline
$A_1:  \left(\begin{array}{ccc}
0&0&\times\\
0&\times&\times\\
 \times&\times&\times
  \end{array}\right)$
& ~$U(1)_{B-2L_{e}-L_{\tau}}$ &
$A_2:  \left(\begin{array}{ccc}
0&\times&0\\
\times&\times&\times\\
 0&\times&\times
  \end{array}\right) $
  &~$U(1)_{B-2L_{e}-L_{\mu}}$ & {Allowed}
\\ \hline
$B_3:  \left(\begin{array}{ccc}
\times&0&\times\\
0&0&\times\\
 \times&\times&\times
  \end{array}\right) $
  &~$U(1)_{B-2L_{\mu}-L_{\tau}}$ &
$B_4:  \left(\begin{array}{ccc}
\times&\times&0\\
\times&\times&\times\\
 0&\times&0
  \end{array}\right) $
  &~$U(1)_{B-2L_{\tau}-L_{\mu}}$ & { Marginally Allowed}
  \\\hline
$D_1:  \left(\begin{array}{ccc}
\times&\times&\times\\
\times&0&0\\
\times&0&\times
  \end{array}\right) $
  &~$U(1)_{B-2L_{\mu}-L_{e}}$ &
$D_2:  \left(\begin{array}{ccc}
\times&\times&\times\\
\times&\times&0\\
\times&0&0
  \end{array}\right) $
  &~$U(1)_{B-2L_{\tau}-L_{e}}$ & { Excluded}
\\\hline \hline
\end{tabular}
\caption{Two texture-zeros and corresponding $U(1)_{B-2L_\alpha-L_\beta}$ symmetry. Here, $\times$ denotes a nonzero matrix element.}
\label{TB:Tex}
\end{table}
\end{center}

Provided all the element in $M_N$ to be equal, then from the texture structure in Eq.\eqref{Eq:hfy} and using Eq.\eqref{mv} we have the $M_\nu$ as
\begin{align}\label{Eq:mv1}
M_\nu\propto\left(
\begin{array}{ccc}
0 & 0 & f_{e2} f_{\tau1}\\
0 & h_{\mu1}^2 & h_{\mu1} h_{\tau2}  \\
f_{e2} f_{\tau1} & h_{\mu1} h_{\tau2} & f_{\tau 1}^2
\end{array}\right),
\end{align}
which is texture $A_1$ allowed by experimental data \cite{Fritzsch:2011qv,Alcaide:2018vni}. Other possible realizations with $U(1)_{B-2L_\alpha-L_\beta}$ can then be easily obtained in a similar approach.
In Tab.~\ref{TB:Tex}, we summarize all the six textures realised by $U(1)_{B-2L_\alpha-L_\beta}$ in our approach. According to Ref.~\cite{Alcaide:2018vni}, texture $A_1$ and $A_2$ predict $\sum m_i\sim 0.07$ eV, hence are allowed by Planck limit $\sum m_i<0.12$ eV \cite{Aghanim:2018eyx}.
Texture $B_3$ and $B_4$ predict $\sum m_i \gtrsim 0.15$ eV, thus are marginally allowed if certain mechanism is introduced to modify cosmology data. Texture $D_1$ and $D_2$ are already excluded by neutrino oscillation data. Following phenomenological predictions are based on texture $A_1$ with $U(1)_{B-2L_e-L_\tau}$.

In the mass eigenstate of heavy Majorana fermion $N_i$, the corresponding Yukawa couplings with leptons are easily obtained by
\begin{align}\label{Eq:hf}
h^\prime=hV=\left(
\begin{array}{cc}
0 & 0\\
h_{\mu 1} V_{11} & h_{\mu 1} V_{12} \\
h_{\tau 2} V_{21} & h_{\tau 2}  V_{22}
\end{array}\right),\quad
f^\prime=fV=\left(
\begin{array}{cc}
f_{e2} V_{21} & f_{e2} V_{22}\\
0 & 0\\
f_{\tau 1} V_{11} & f_{\tau1} V_{12}
\end{array}\right).
\end{align}
For the $Z_2$-even scalars, the CP-even scalars in weak-basis ($\sqrt{2}\Re \Phi^0, \sqrt{2}\Re S_1, \sqrt{2}\Re S_2$) mix into mass-basis ($h,H_1, H_2$) with mass spectrum $M_h\sim M_{H_1}<M_{H_2}$. Without loss of generality, we further assume mixing angle between ($h,H_1$) being $\alpha$ and vanishing mixing angles between $H_2$ and $h/H_1$ for simplicity. The would-be Goldstone boson $\Phi^+,\sqrt{2}\Im \Phi^0,\sqrt{2}\Im S_2$ are absorbed by gauge boson $W^+,Z,Z'$ respectively, leaving $\sqrt{2}\Im S_1$ a massless Mojoron $J$. In principle, if we introduce $U(1)_D$ gauge symmetry to produce the discrete $Z_2$ symmetry, this Mojoron $J$ could be absorbed by the dark gauge boson $Z_D$ \cite{Ma:2013yga}. For $Z_2$-odd scalars, there is no mixing between $\eta_1$ and $\eta_2$. Since texture of $M_\nu$ in Eq. \eqref{scale} is derived by $m_0^2\gg M_{Nk}^2$, only fermion DM is allowed in this paper.

\section{Phenomenology}\label{Sec:PH}

\subsection{Neutrino Mixing}

\begin{figure}
\begin{center}
\includegraphics[width=0.443\linewidth]{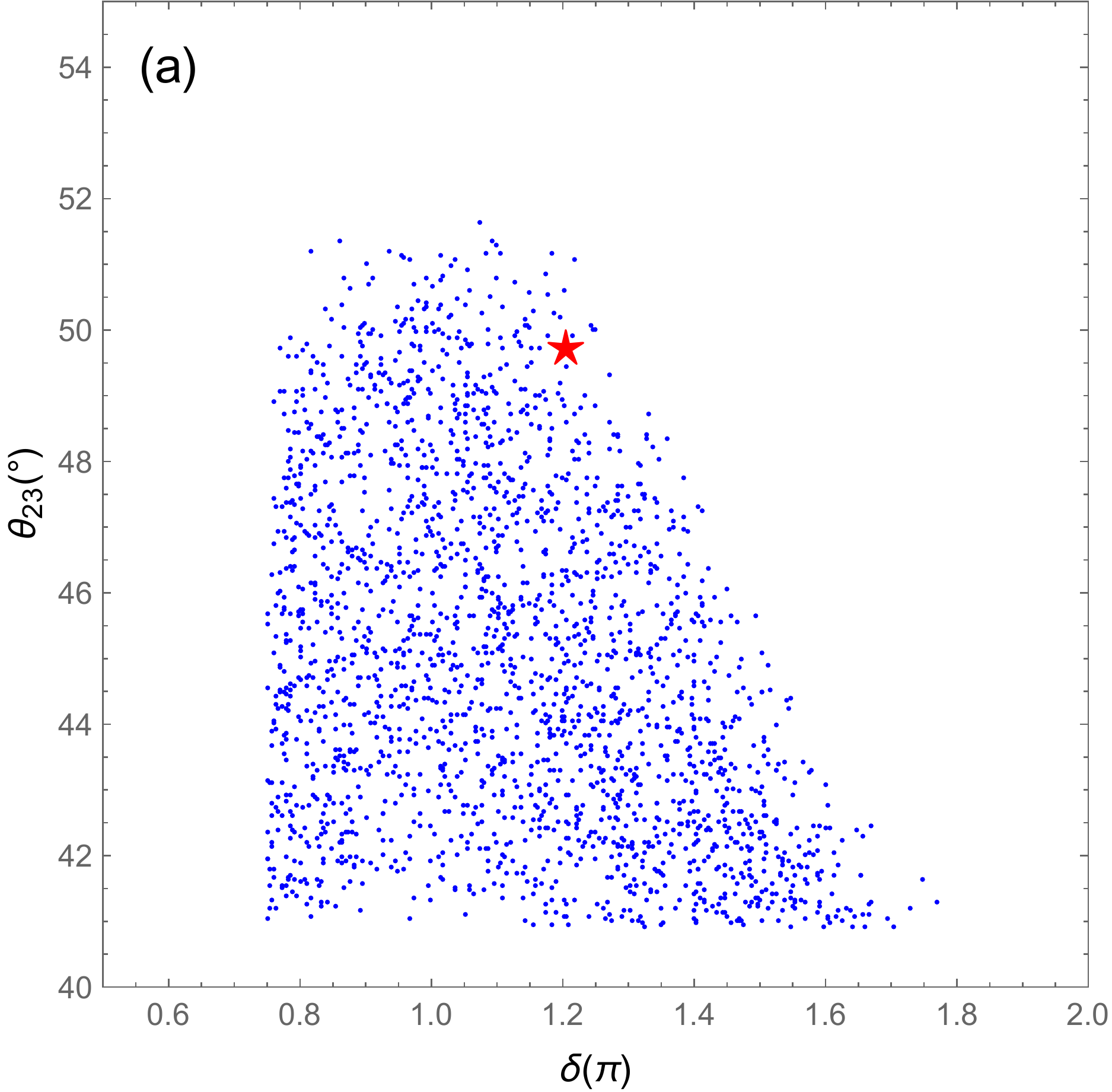}
\includegraphics[width=0.45\linewidth]{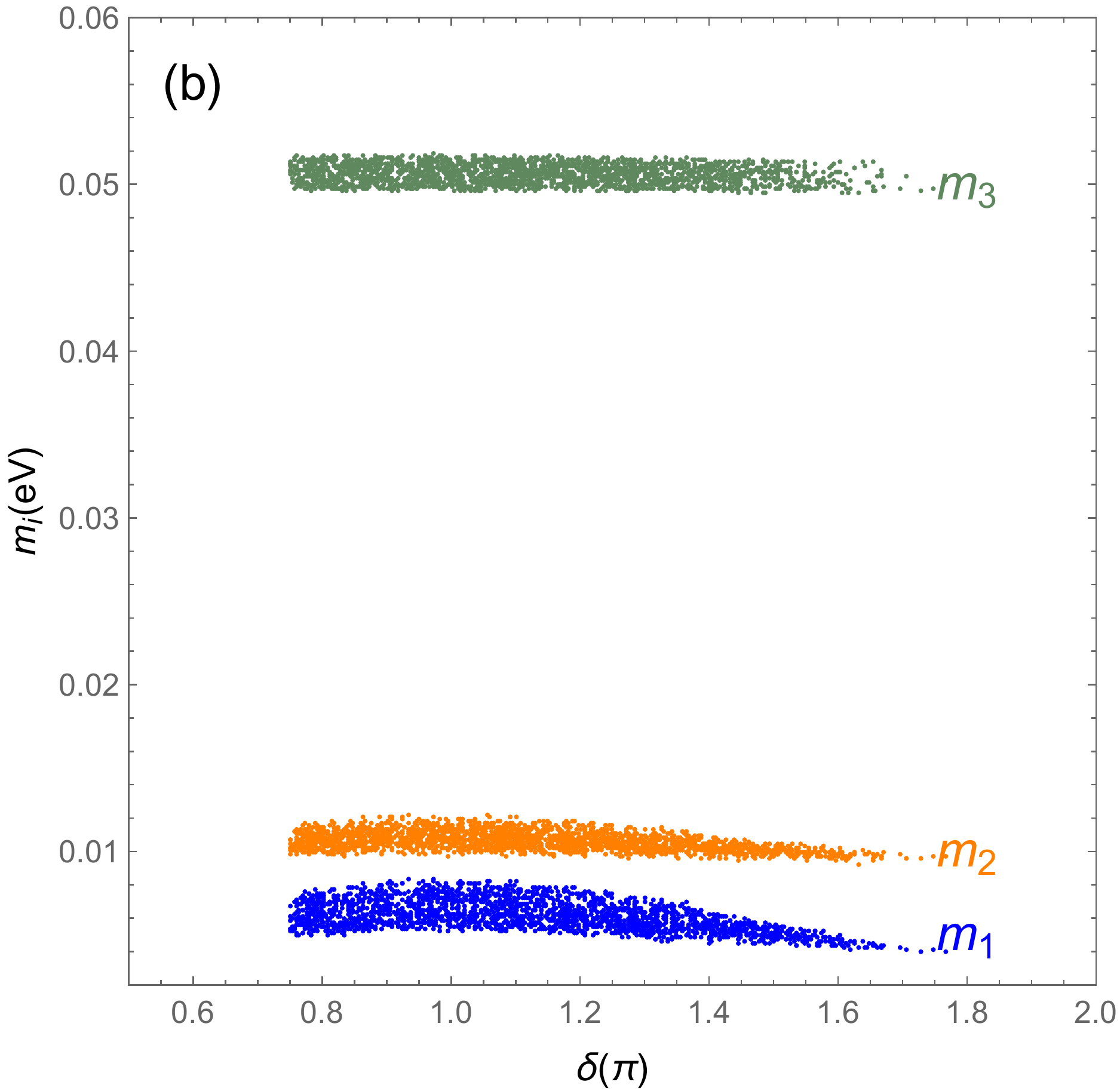}\\
\includegraphics[width=0.45\linewidth]{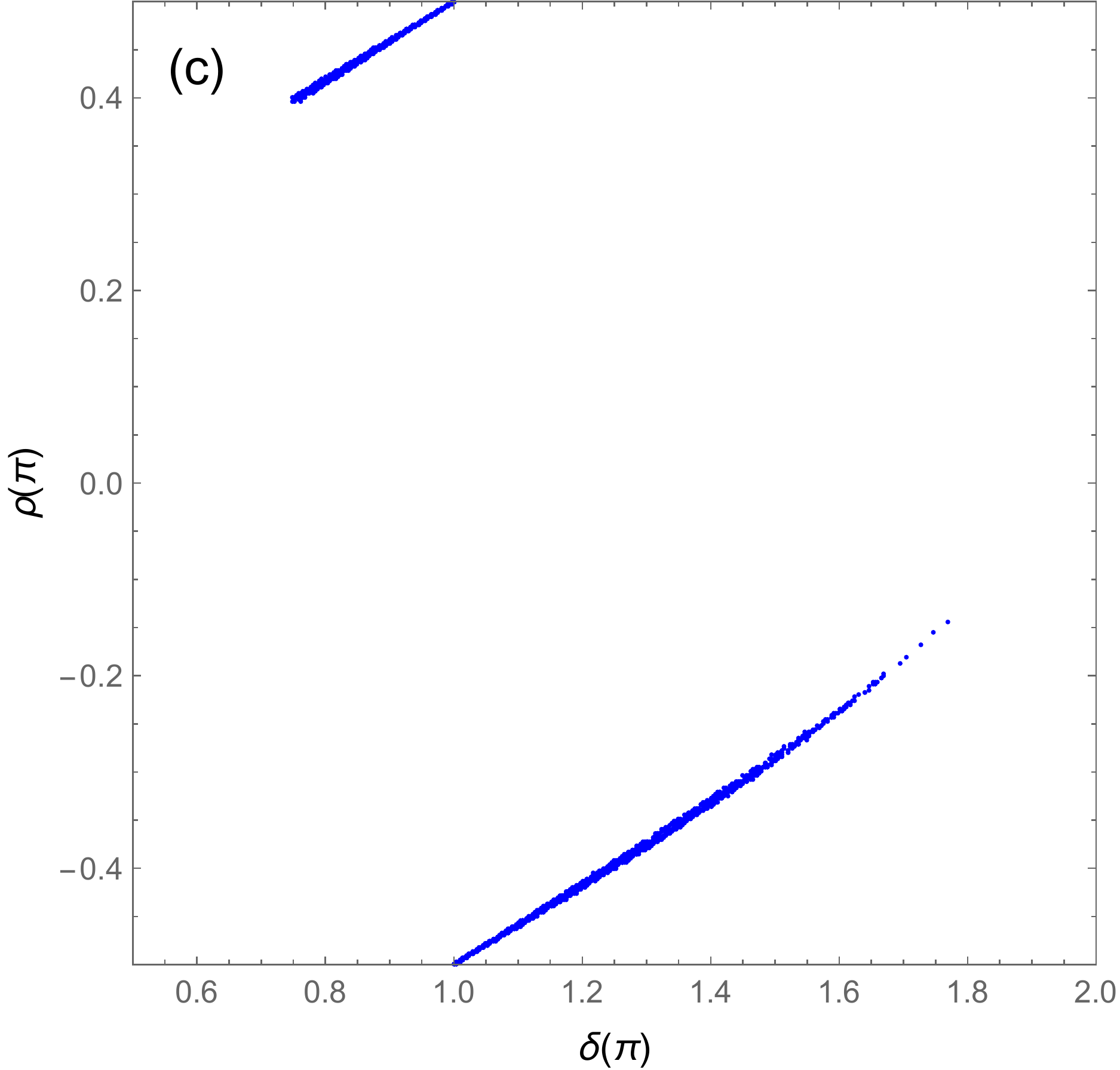}
\includegraphics[width=0.45\linewidth]{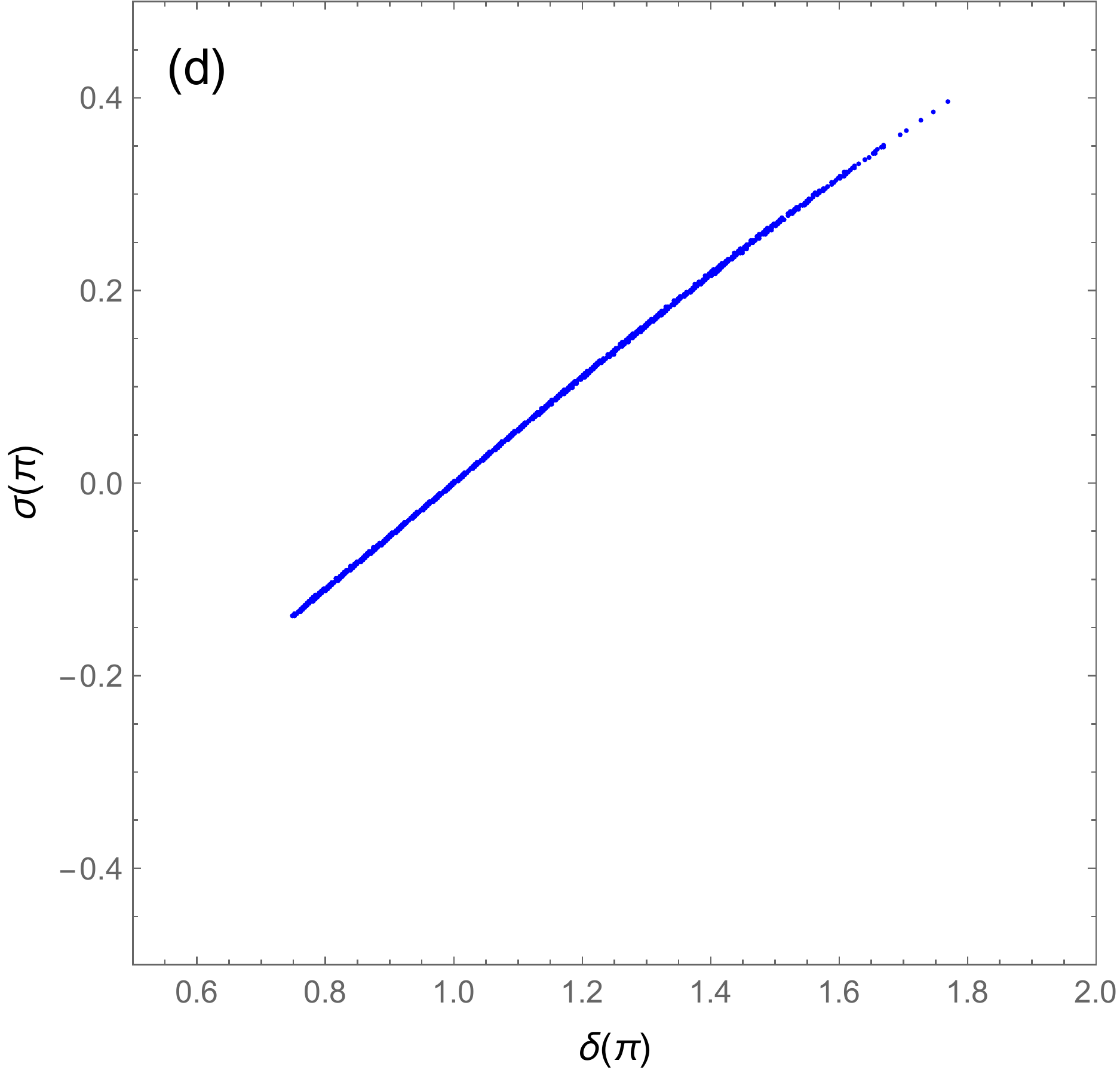}
\end{center}
\caption{Allowed samples of $A_1$ texture with neutrino oscillation data varied in $3\sigma$ range of Ref.~\cite{Esteban:2018azc}. In pattern (a), the red star {\color{red} $\bigstar$} stands for the best fit point from global analysis.
\label{Fig:TZ}}
\end{figure}

The flavor dependent $U(1)_{B-2L_e-L_\tau}$ symmetry leads to texture $A_1$ \cite{Fritzsch:2011qv,Alcaide:2018vni}. Since $(M_\nu)_{ee}=0$, the predicted effective Majorana neutrino mass $\langle m \rangle_{ee}$ is exactly zero for the neutrinoless double-beta decay. Therefore, only normal hierarchy is allowed \cite{Bilenky:1999wz,Vissani:1999tu}. Following the procedure in Ref.~\cite{Fritzsch:2011qv}, we now update the predictions of neutrino oscillation data with latest global analysis results \cite{Esteban:2018azc}.

In Fig.~\ref{Fig:TZ}, we show the scanning results of texture $A_1$.
It is worth to note that the best fit value of neutrino oscillation parameters by global analysis \cite{Esteban:2018azc} is only marginally consistent with predictions of texture $A_1$, which is clearly seen in Fig.~\ref{Fig:TZ}~(a). From Fig.~\ref{Fig:TZ}~(b), we obtain that $m_1\sim 0.007$ eV, $m_2\sim 0.01$ eV, and $m_3\approx\sqrt{\Delta m^2}\sim 0.05$ eV. The resulting sum of neutrino mass is then $\sum m_i\sim 0.07$ eV, thus it satisfies the bound from cosmology, i.e., $\sum m_i<0.12$ eV \cite{Aghanim:2018eyx}. The Dirac phase should fall in the range $\delta\in[0.75\pi, 1.77\pi]$, meanwhile Fig.~\ref{Fig:TZ} (c) and (d) indicate that $\rho\approx\frac{\delta}{2}$ and $\sigma\approx\frac{\delta}{2}-\frac{\pi}{2}$.

Instead of the marginally best fit value, we take $\delta=\pi$ and $\theta_{23}=46^\circ$ with other oscillation parameters being the best fit value in Ref~\cite{Esteban:2018azc} as the benchmark point for illustration, which leads to the following neutrino mass structure
\begin{align}\label{Eq:mv2}
M_\nu=\left(
\begin{array}{ccc}
0 & 0 & 0.0110\\
0 & 0.0293 & 0.0219  \\
0.0110 & 0.0219 & 0.0256
\end{array}\right)~\text{eV}
\end{align}
By comparing the analytic $M_\nu$ in Eq.~\eqref{Eq:mv1} and numerical $M_\nu$ in Eq.~\eqref{Eq:mv2}, one can easily reproduce the observed neutrino oscillation data by requiring
\begin{eqnarray}\label{Eq:BF}
\frac{h_{\tau 2}}{h_{\mu1}}:\frac{f_{\tau1}}{h_{\mu1}}:\frac{f_{e2}}{h_{\mu1}}
 &=& \frac{(M_\nu)_{\mu\tau}}{(M_\nu)_{\mu\mu}}:
 \sqrt{\frac{(M_\nu)_{\tau\tau}}{(M_\nu)_{\mu\mu}}}:
 \frac{(M_\nu)_{e\tau}}{\sqrt{(M_\nu)_{\mu\mu}(M_\nu)_{\tau\tau}}}\\ \nonumber
 & \simeq &0.745:0.933:0.401.
\end{eqnarray}
Hence, we can take $h_{\mu1}$ as free parameters and determine the other three Yukawa coupling by using above ratios. The overall neutrino mass scale is then determined by $\lambda v^2 M_N h_{\mu1}^2/(32\pi^2 m_0^2)\approx 0.0293$ eV.

\begin{figure}
\begin{center}
\includegraphics[width=0.45\linewidth]{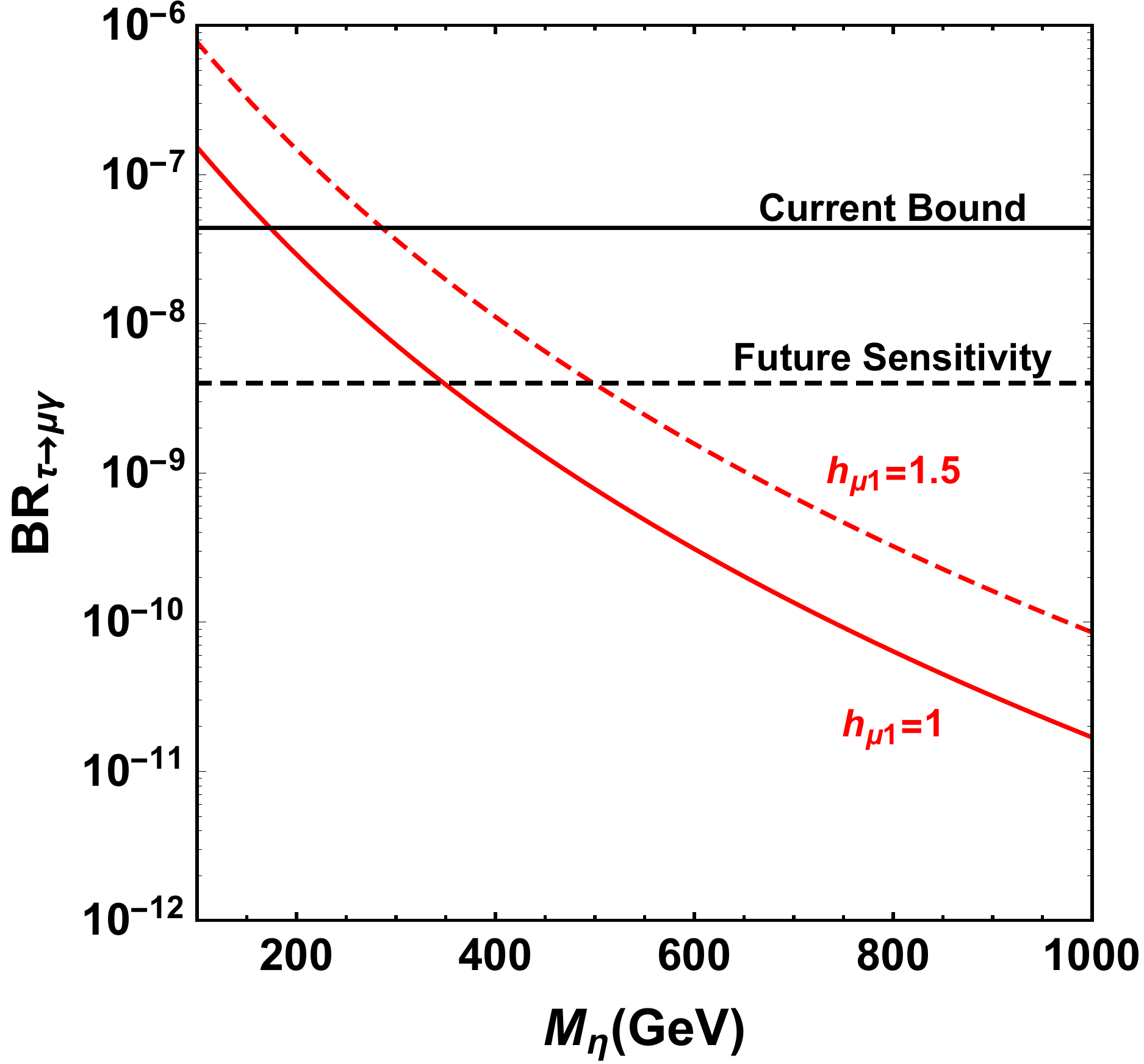}
\includegraphics[width=0.45\linewidth]{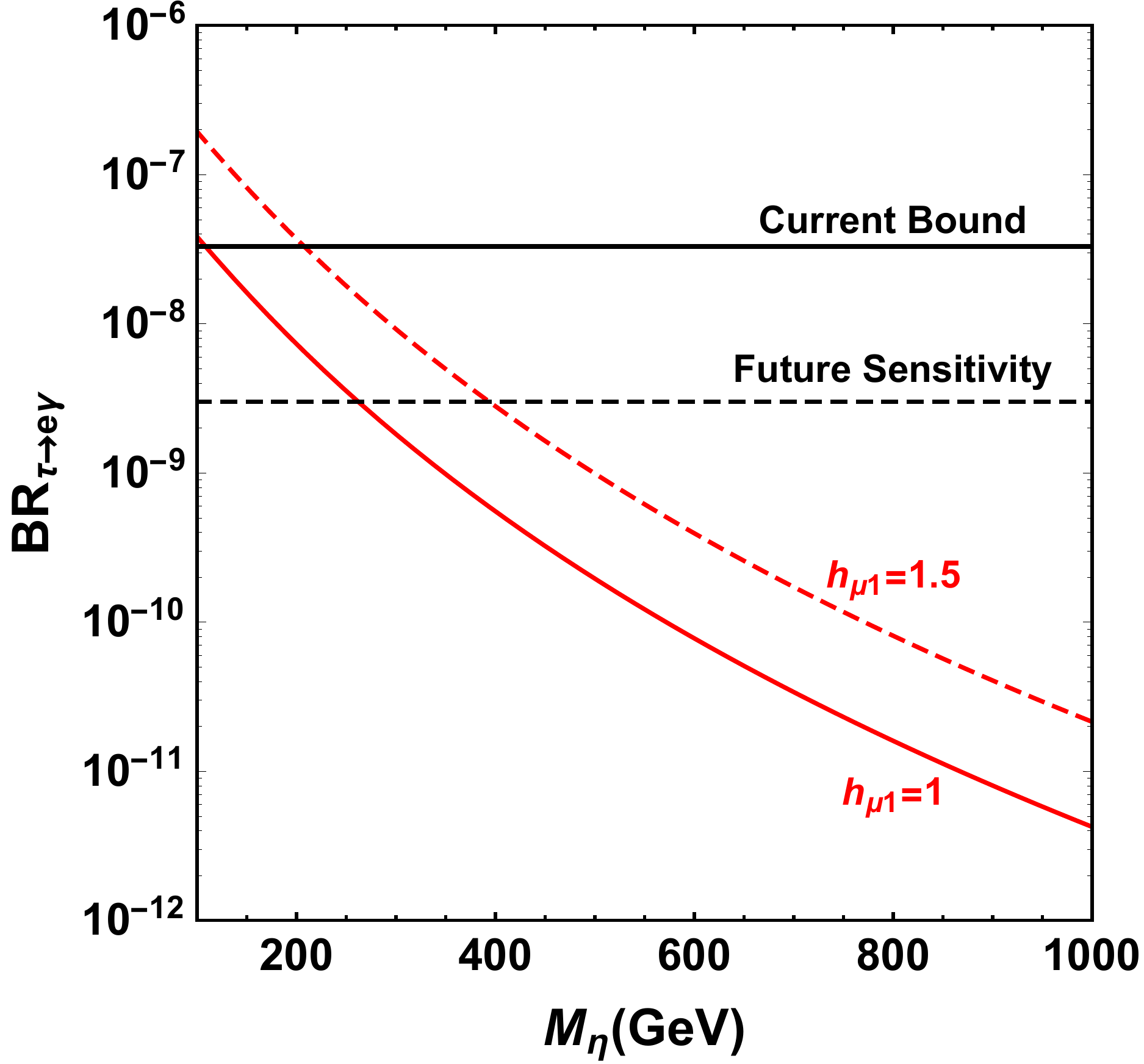}
\end{center}
\caption{Predictions for $\tau\to \mu\gamma$ (left) and $\tau \to e\gamma$ (right) with corresponding current bound \cite{Aubert:2009ag} and future sensitivity \cite{Hayasaka:2013dsa}. In these figures, we have fixed $M_{N_1}=200$ GeV.
\label{Fig:LFV}}
\end{figure}

\subsection{Lepton Flavor Violation}

The new Yukawa interactions of the form $\bar{L}\tilde{\eta} N_R$ will contribute to lepton flavor violation (LFV) processes \cite{Toma:2013zsa,Ding:2014nga}. In this work, we take the radiative decay $\ell_i\to \ell_j \gamma$ for illustration. With flavor dependent $U(1)_{B-2L_e-L_\tau}$ symmetry, it is clear from Eq.~\eqref{Eq:Yuk} that $\eta_1^\pm(\eta_2^\pm)$ will only induce $\tau\to\mu\gamma(\tau\to e\gamma)$ at one-loop level. It is worth to note that the most stringent $\mu\to e\gamma$ decay is missing at one-loop level. Hence, if the ongoing experiments observe $\tau\to\mu\gamma$ and $\tau\to e\gamma$ but no $\mu\to e \gamma$, this model will be favored. The corresponding branching ratios are calculated as
\begin{eqnarray}
\text{BR}(\tau\to \mu\gamma) &=& \frac{3\alpha}{64\pi G_F^2}
\left|\sum_{i=1}^2 \frac{(h_{\mu1} V_{1i})(h_{\tau2} V_{2i})^*}{M_{\eta_1}^2}
F\left(\frac{M_{Ni}^2}{M_{\eta_1}^2}\right)\right|^2
\text{BR}(\tau\to \mu \nu_\tau\bar{\nu}_\mu), \\ \nonumber
\text{BR}(\tau\to e\gamma) &=& \frac{3\alpha}{64\pi G_F^2}
\left|\sum_{i=1}^2 \frac{(f_{e2} V_{2i})(f_{\tau1} V_{1i})^*}{M_{\eta_2}^2}
F\left(\frac{M_{Ni}^2}{M_{\eta_2}^2}\right)\right|^2
\text{BR}(\tau\to e \nu_\tau\bar{\nu}_e),
\end{eqnarray}
where the loop function $F(x)$ is
\begin{equation}
F(x)=\frac{1-6x+3x^2+2x^3-6x^2\ln x}{6(1-x)^4}.
\end{equation}
In the limit for degenerate $M_N$, we have
\begin{equation}
\text{BR}(\tau\to \ell \gamma) \propto \left|\sum_{i=1}^2 V_{1i} V_{2i}^*\right|^2=|(VV^\dag)_{12}|^2 =0,
\end{equation}
where in the last step, we have considered the fact that $V$ is an unitary matrix. Therefore, large cancellations between the contribution of two $N_i$ are also possible even in the case of non-degenerate $M_N$. In Fig.~\ref{Fig:LFV}, we show the predictions for $\tau\to \mu\gamma$ and $\tau\to e \gamma$. Although constraint on BR($\tau\to e\gamma$) is slightly more stringent than BR($\tau\to \mu\gamma$), the predicted BR($\tau\to e\gamma$) is much smaller than BR($\tau\to \mu\gamma$). It is clear that the current bound is quite loose, e.g., $M_\eta\gtrsim200$ GeV with $h_{\mu 1}=1$ can be allowed.

\begin{figure}
\begin{center}
\includegraphics[width=0.45\linewidth]{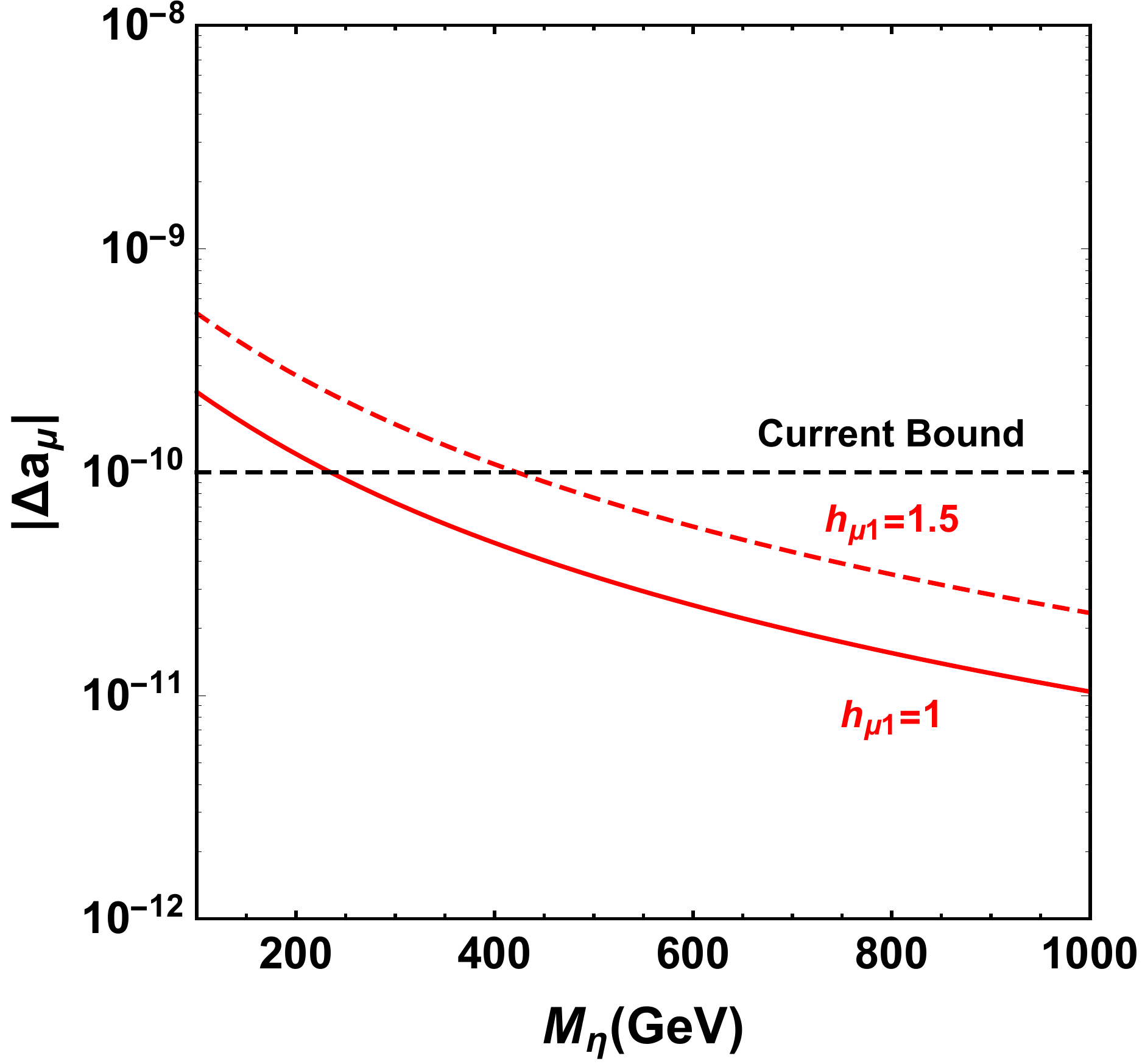}
\end{center}
\caption{Predictions for $|\Delta a_\mu|$. In this figures, we have fix $M_{N_1}=200$ GeV.
\label{Fig:g-2}}
\end{figure}

Although the Yukawa interaction $\bar{L}_\mu\tilde{\eta}_1 N_{i}$ can not induce $\mu \to e\gamma$ at one-loop, it does contribute to muon anomalous magnetic moment \cite{Lindner:2016bgg}
\begin{equation}
\Delta a_\mu = - \sum_{i=1}^2 \frac{  |h_{\mu1} V_{1i}|^2 M_\mu^2}{16\pi^2 M_{\eta_1}^2} F\left(\frac{M_{Ni}^2}{M_{\eta_1}^2}\right).
\end{equation}
Comparing with BR($\tau\to \ell\gamma$), there is no cancellations between the contribution of two $N_i$. However, the total contribution to $\Delta a_\mu$ is negative, while the observed discrepancy $\Delta a_\mu= a_\mu^\text{EXP}-a_\mu^\text{SM}=(261\pm78)\times10^{-11}$ is positive \cite{Blum:2013xva}. Thus, the Yukawa interaction $\bar{L}_\mu\tilde{\eta}_1 N_{R1}$ can not explain the $(g-2)_\mu$ anomaly, and some other new physics is required \cite{Lindner:2016bgg}. On the other hand, since a too large negative contribution to $\Delta a_\mu$ is not favored, we consider theoretical $|\Delta a_\mu|<10^{-10}$ in the following. The results are shown in Fig.~\ref{Fig:g-2}. We find that the bound from $|\Delta a_\mu|<10^{-10}$ is actually slightly more stringent than BR$(\tau\to \mu\gamma)$.

\subsection{Dark Matter}

\begin{figure}
\begin{center}
\includegraphics[width=0.45\linewidth]{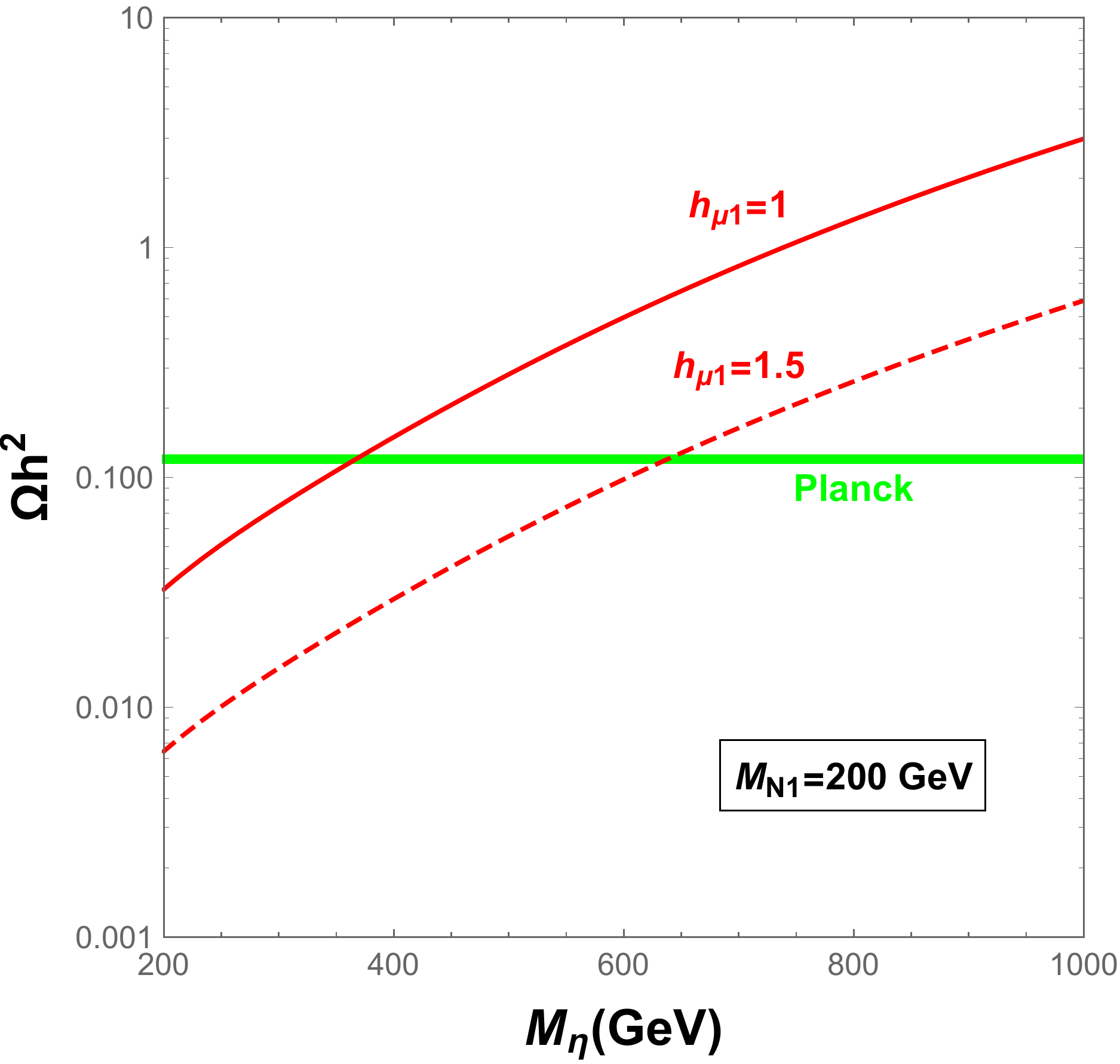}
\end{center}
\caption{Predicted relic density as a function of $m_\eta$, where we have fix $M_{N_1}=200$ GeV. The green line corresponds to the observed relic density $\Omega h^2=0.120\pm0.001$ \cite{Aghanim:2018eyx}.
\label{Fig:RD}}
\end{figure}

In this work, we consider $N_1$ is the DM candidate. In the original scotogenic model \cite{Ma:2006km}, the viable annihilation channel is $N_1 N_1 \to \ell^+\ell^-,\bar{\nu}\nu$ via the Yukawa interaction $\bar{L}_\ell\tilde{\eta} N_{1}$ \cite{Kubo:2006yx}. However, such annihilation channel is tightly constrained by non-observation of LFV \cite{Vicente:2014wga}. Thanks to relative loose constraints from $\tau$ decays, the scanning results of Ref.~\cite{Vicente:2014wga} suggested that $N_1$ should have a large coupling to $L_\tau$. Thus, the dominant annihilation channel is $\tau^+\tau^-$ and $\bar{\nu}_\tau\nu_\tau$ with $M_{N_1}\lesssim 3$ TeV.

Quite different from the original scotogenic model \cite{Ma:2006km}, the LFV process is either vanishing or suppressed in this flavor dependent model. Therefore, $\mathcal{O}(1)$ Yukawa coupling can be easily realised without tuning. In the following quantitative investigation, we consider a special scenario, i.e., $M_{\eta_1}=M_{\eta_2}=M_\eta$ for simplicity. For vanishing lepton masses, the Yukawa-portal annihilation cross section is \cite{Kubo:2006yx,Li:2010rb}
\begin{equation}
\sigma v_\text{rel} = a + b v_\text{rel}^2 = 0+
\sum_{\alpha,\beta}\left|h^\prime_{\alpha1}h^{\prime*}_{\beta1}
+f^\prime_{\alpha1}f^{\prime*}_{\beta1}\right|^2
 \frac{r^2(1-2r+2r^2)}{24\pi M_{N_1}^2} v_\text{rel}^2,
\end{equation}
where $v_\text{rel}$ is the relative speed, $h^\prime$ and $f^\prime$ are defined in Eq.~\eqref{Eq:hf}, $r=M_{N_1}^2/(M_{\eta}^2+M_{N_1}^2)$.  The thermally averaged cross section is calculated as $\langle \sigma v_\text{rel}\rangle= a + 6 b/x_f$,  where the freeze-out parameter $x_f=M_{N_1}/T_f$ is obtained by numerically solving
\begin{equation}
x_f = \ln \left(\frac{0.038 M_{\text{Pl}} M_{N_1} \langle \sigma v_{\text{rel}}\rangle}{\sqrt{g_* x_f}}\right).
\end{equation}
The relic density is then calculated as \cite{Bertone:2004pz}
\begin{equation}
\Omega h^2 = \frac{1.07\times10^9 \GeV^{-1}}{ M_{\text{Pl}}}
\frac{x_f}{\sqrt{g_*}}\frac{1}{a+3b/x_f},
\end{equation}
where $M_\text{Pl}=1.22\times10^{19}$ GeV is the Planck mass, $g_*$ is the number of relativistic degrees of freedom. The numerical results are depicted in Fig~\ref{Fig:RD}. Provided the mass of the DM candidate is $M_{N_1}=200$ GeV, then the observed relic density is interpreted by $h_{\mu1}=1, M_\eta=366$ GeV or $h_{\mu1}=1.5, M_\eta=640$ GeV. That is to say, $h_{\mu1}\sim\mathcal{O}(1)$ is required to obtain correct relic density, and the larger $h_{\mu1}$ is, the larger the mass splitting $M_\eta-M_{N_1}$ is.

In addition to the Yukawa-portal interaction, $N_1$ can also annihilate via the Higgs-portal and $Z'$-portal interactions \cite{Okada:2010wd,Kanemura:2011vm,
Okada:2012sg,Wang:2015saa,Okada:2016gsh,Okada:2018ktp,Han:2018zcn}.
In these two scenarios, $M_{N_1}\simeq M_{h,H}/2$ or $\simeq M_{Z'}/2$ are usually required to realize correct relic density \cite{Borah:2018smz}. If the additional scalar singlet scalar $H$ is lighter than $N_1$, then the annihilation channel $N_1 N_1\to HH$ with $H\to b\bar{b}$ is able to explain the Fermi-LAT gamma-ray excess at the Galactic center \cite{Kim:2016csm,Ding:2018jdk}.

\begin{figure}
\begin{center}
\includegraphics[width=0.445\linewidth]{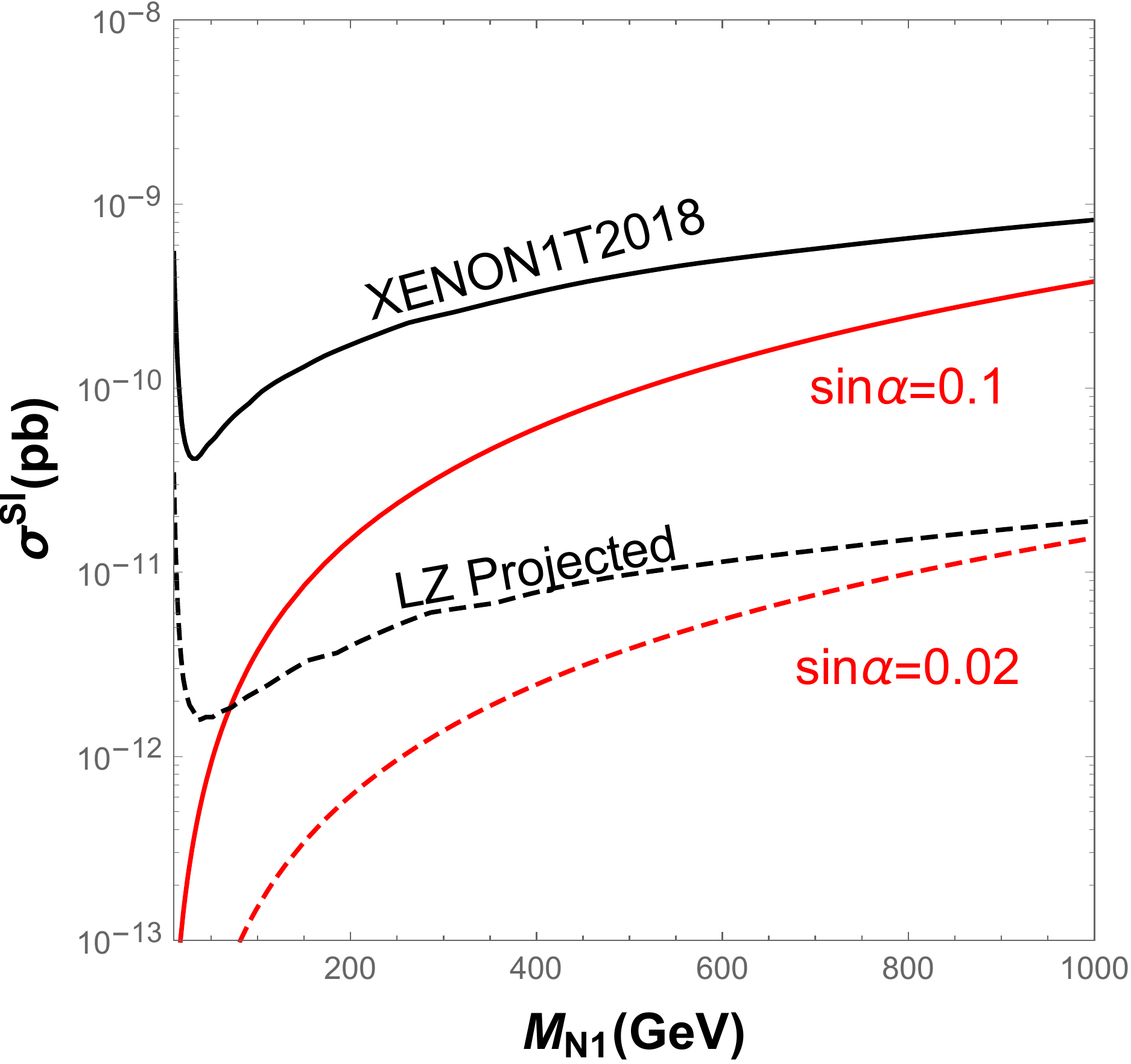}
\end{center}
\caption{Spin-independent cross section as a function of $M_{N_1}$. The black solid and dashed line correspond to current XENON1T \cite{Aprile:2018dbl} and future LZ \cite{Akerib:2018lyp} limits, respectively. In this figure, we have set $M_{H_1}=500$ GeV and $v_S=10$ TeV.
\label{Fig:DD}}
\end{figure}

The spin-independent DM-nucleon scattering cross section is dominantly mediated by scalar interactions, which is given by
\begin{equation}
\sigma^\text{SI}=\frac{4}{\pi}\left(\frac{M_p M_{N_1}}{M_p+M_{N_1}}\right)^2 f_p^2,
\end{equation}
where $M_p$ is the proton mass and the hadronic matrix element $f_p$ reads
\begin{equation}
\frac{f_p}{M_p}=\sum_{q=u,d,s}f^p_{Tq} \frac{\alpha_q}{M_q}+
\frac{2}{27}\left(1-\sum_{q=u,d,s}f^p_{Tq}\right)\sum_{q=c,b,t}\frac{\alpha_q}{M_q}.
\end{equation}
and the effective vertex
\begin{equation}
\frac{\alpha_q}{M_q}=-\frac{y_{N_1}}{\sqrt{2}v}\sin2\alpha
\left(\frac{1}{M_h^2}-\frac{1}{M_{H_1}^2}\right),
\end{equation}
Here, $y_{N_1}=y_{11} V_{11}^2$ is the effective Yukawa coupling of $N_1$ with $S_1$. For proton, the parameters $f^p_{Tq}$ are evaluated as $f^p_{Tu}=0.020\pm0.004$, $f^p_{Td}=0.026\pm0.005$ and $f^p_{Ts}=0.118\pm0.062$ \cite{Ellis:2000ds}. Fig.~\ref{Fig:DD} shows the numerical results for direct detection. It is obvious that the predicted $\sigma^\text{SI}$ with $\sin\alpha=0.1$ lies below current XENON1T limit, but the range of $M_{N_1}\gtrsim70$ GeV is within future LZ's reach. However, if no direct detection signal is observed by LZ, then $\sin\alpha\lesssim0.02$ should be satisfied.

\begin{figure}
\begin{center}
\includegraphics[width=0.445\linewidth]{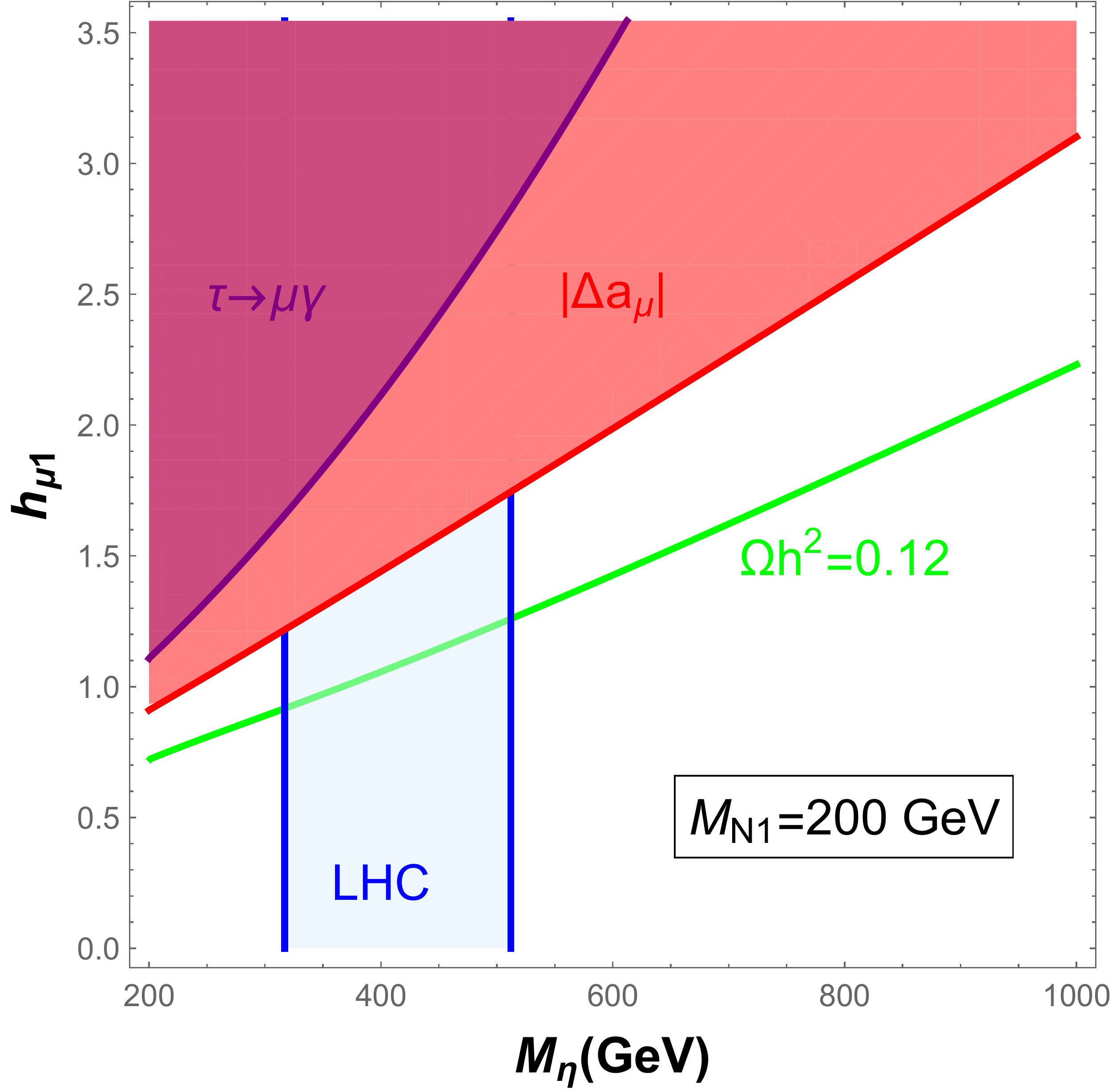}
\includegraphics[width=0.45\linewidth]{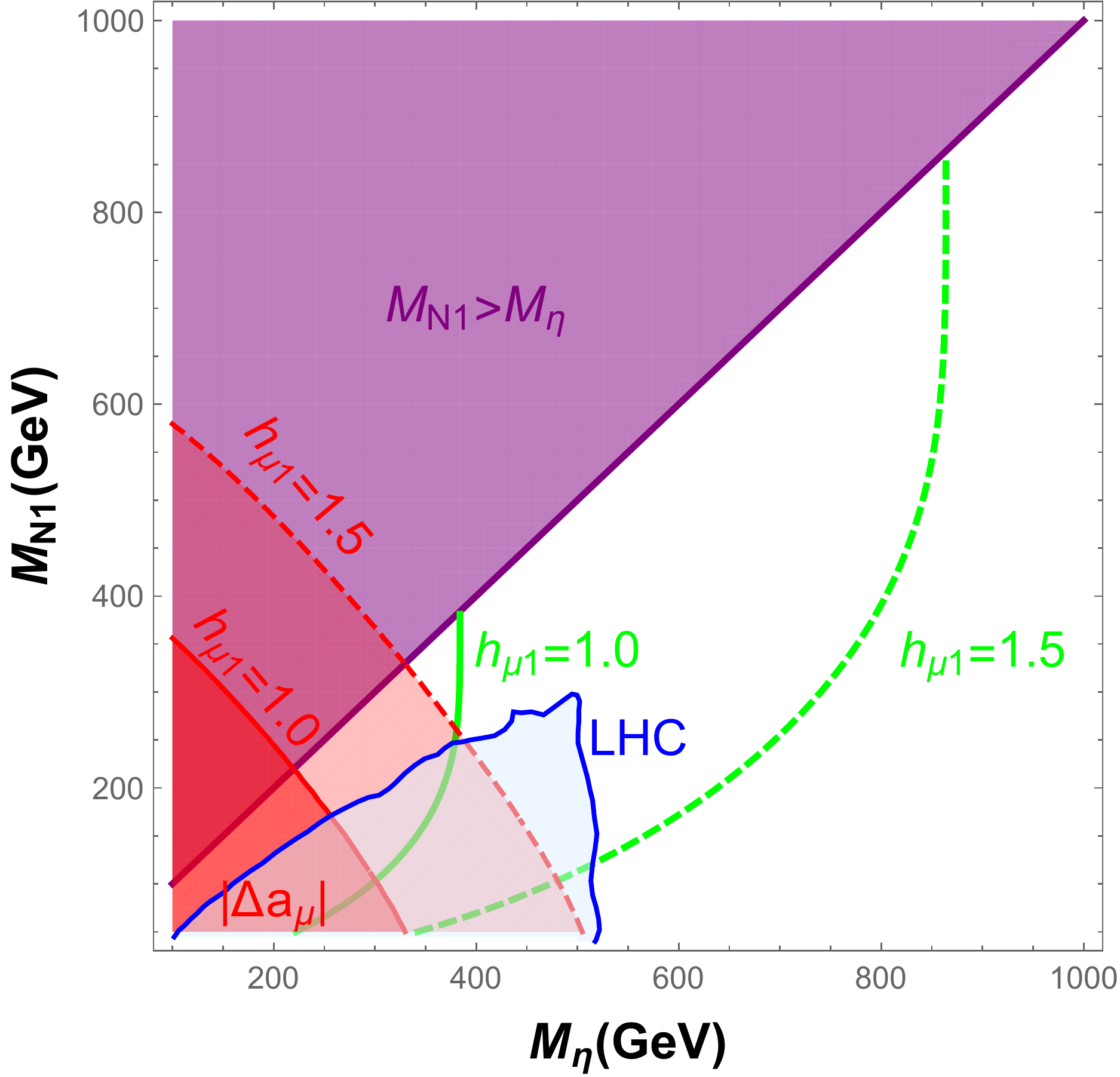}
\end{center}
\caption{ Combined results for the Yukawa-portal DM. Left pattern: in the $h_{\mu1}$-$M_\eta$ plane; right pattern: in the $M_{N_1}$-$M_\eta$ plane. The green lines satisfy the condition for correct relic density. And the blue regions are excluded by LHC direct search, which will be discussed in Sec.~\ref{Sec:CS}.
\label{Fig:CB}}
\end{figure}

In Fig.~\ref{Fig:CB}, we show the combined results from LFV, $|\Delta a_\mu|$, relic density and LHC search. In left pattern of Fig.~\ref{Fig:CB}, it indicates that for $M_{N_1}=200$ GeV, the only exclusion region is from LHC search. Hence, either $M_\eta\lesssim 300$ GeV with $h_{\mu1}\lesssim0.9$ or $M_\eta\gtrsim500$ GeV with $h_{\mu1}\gtrsim1.3$ is required. In right pattern of Fig.~\ref{Fig:CB}, two benchmark value $h_{\mu1}=1.0,1.5$ are chosen to illustrate. For $h_{\mu1}=1.0$, we have $250~\text{GeV}\lesssim M_{N_1}\lesssim M_\eta\sim 400$ GeV. Therefore, the only viable region is $M_{N_1}\sim M_{\eta}\lesssim400$ GeV for $h_{\mu1}\lesssim1$. Meanwhile for $h_{\mu1}=1.5$, $M_{N_1}\gtrsim120$ GeV with $M_\eta\gtrsim520$ GeV is able to escape LHC limit.
\subsection{Collider Signature}\label{Sec:CS}

In this part, we highlight some interesting collider signatures. Begin with the newly discovered 125 GeV Higgs boson $h$ \cite{Aad:2012tfa,Chatrchyan:2012xdj}.
The existence of massless Mojoron $J$ will induce the invisible decay of SM Higgs
via $h\to JJ$ \cite{Wang:2016vfj}. The corresponding decay width is evaluated as
\begin{equation}
\Gamma(h\to JJ)\simeq \frac{M_h^3 \sin^2\alpha}{32\pi v_S^2}.
\end{equation}
Then, the branching ratio of invisible decay is BR($h\to JJ)=\Gamma(h\to JJ)/(\Gamma(h\to JJ)+\Gamma_\text{SM}\cos^2\alpha)$, where $\Gamma_\text{SM}=4.09$ MeV \cite{deFlorian:2016spz}. Currently, the combined direct and indirect observational limit on invisible Higgs decay is BR($h\to JJ)<0.24$ \cite{Khachatryan:2016whc}. Typically for $\sin\alpha=0.1, v_S=10$ TeV, we have BR($h\to JJ)=4.8\times10^{-4}$, which is far below current limit.
Meanwhile, if $M_{H_1}<2 M_h$, then $h\to H_1H_1$ with $H_1\to JJ$ will also contribute to invisible Higgs decay ~\cite{Bonilla:2015uwa}.

\begin{figure}
\begin{center}
\includegraphics[width=0.45\linewidth]{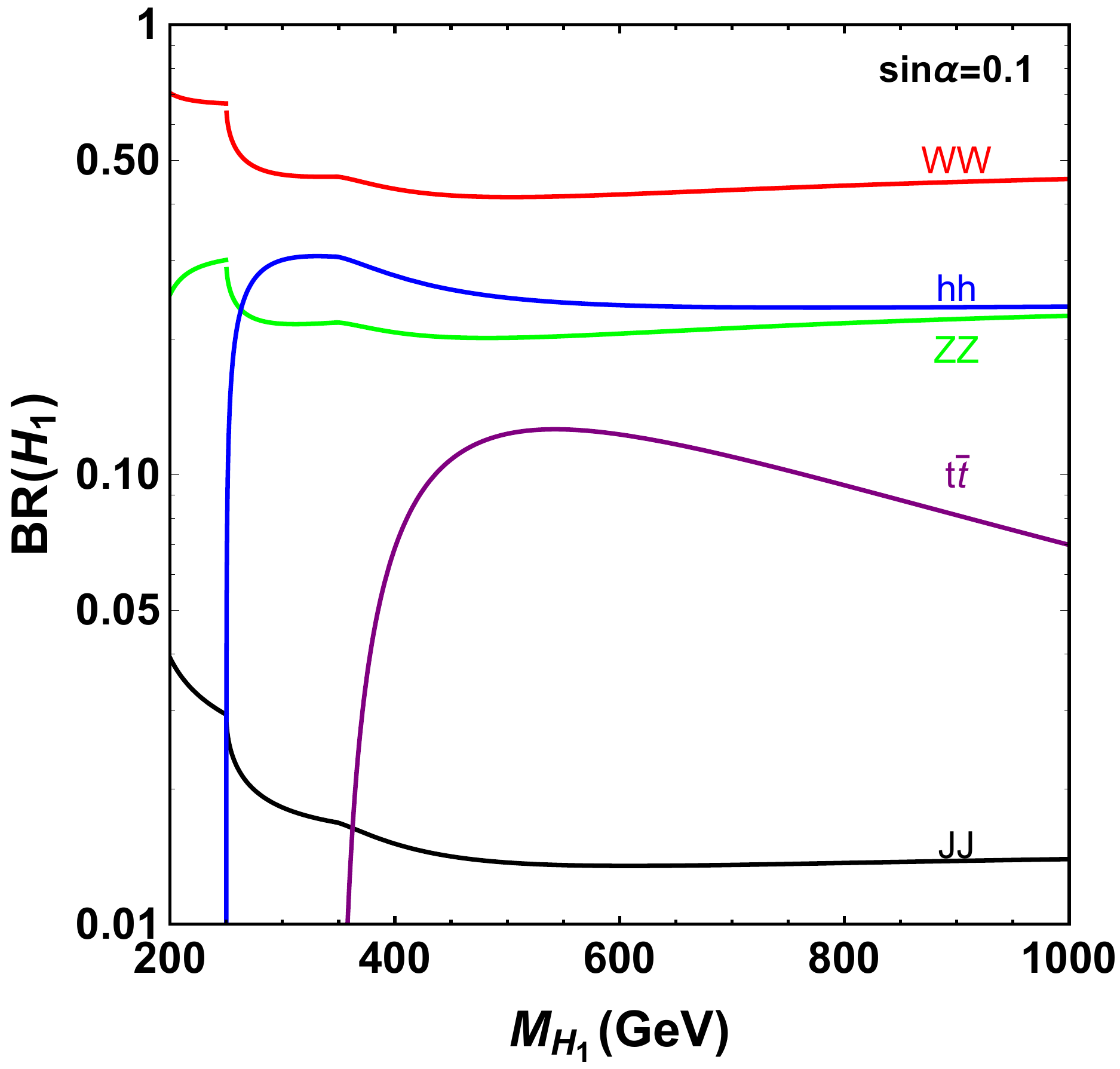}
\includegraphics[width=0.45\linewidth]{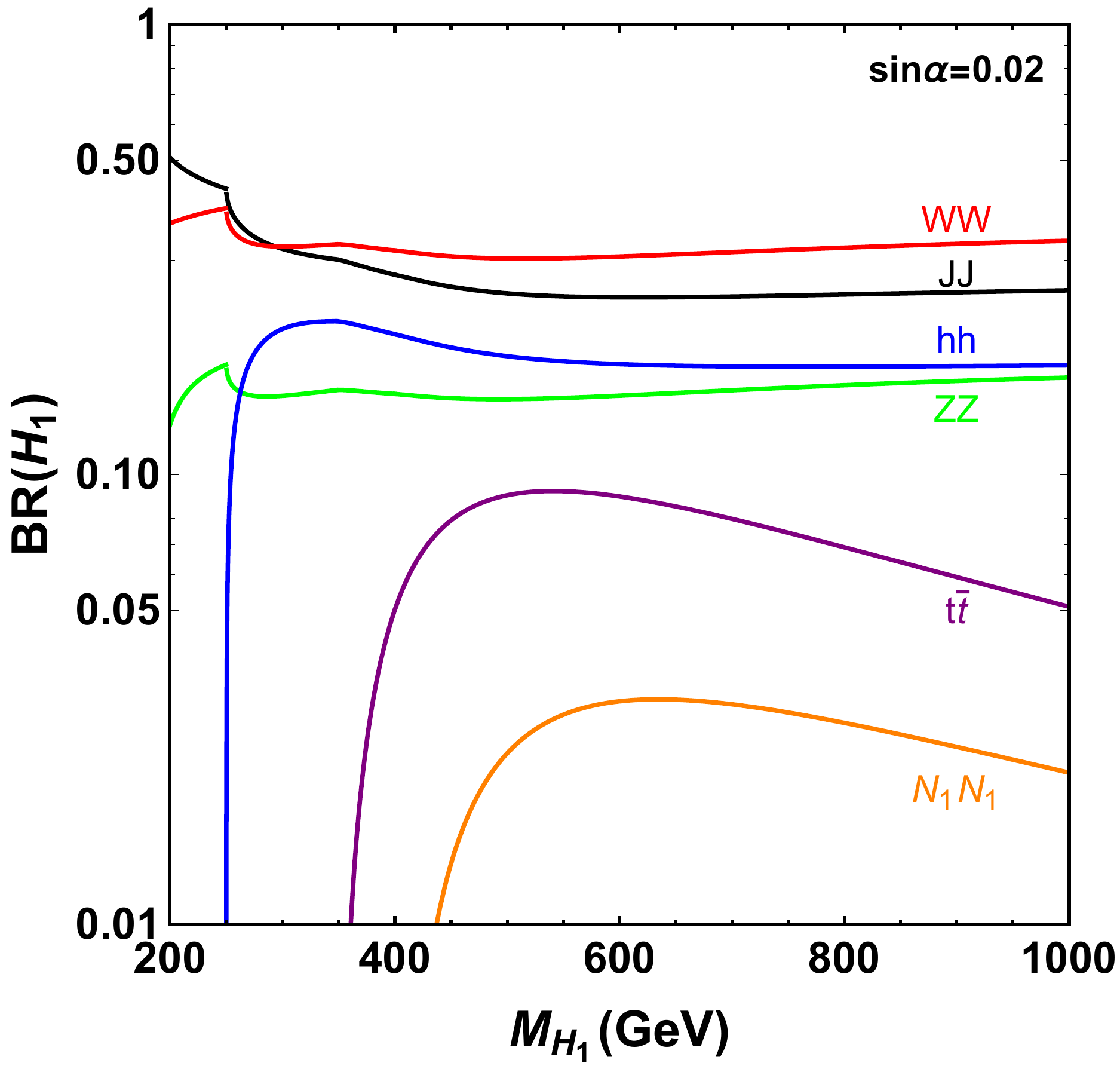}
\end{center}
\caption{ Branching ratios of scalar singlet $H_1$ for $\sin\alpha=0.1$(left) and $\sin\alpha=0.02$(right). In this figures, we have also fix $M_{N_1}=200$ GeV and $v_S=10$ TeV. Note in left pattern, BR$(H_1\to N_1 N_1)$ is less than 0.01, thus is not shown in the plot.
\label{Fig:H1}}
\end{figure}

In this paper, we consider the high mass scenario $M_{H_1}>M_h$. In addition to the usual $H_1\to$ SM final states as real singlet model \cite{Robens:2016xkb}, the heavy scalar singlet can also decay into Majoron pair $H_1\to JJ$ and DM pair $H_1\to N_1 N_1$. Fig.~\ref{Fig:H1} shows the dominant decay branching ratios of $H_1$. The invisible BR$(H_1\to JJ)$ is less than 0.02 when $\sin\alpha=0.1$, therefore $H_1$ appears as a SM heavy Higgs with BR$(H_1\to hh)\approx\text{BR}(h\to ZZ)\approx\frac{1}{2}\text{BR}(H_1\to WW)\approx \frac{1}{4}$. While for $\sin\alpha=0.02$, the invisible BR$(H_1\to JJ)$ increases to about $0.25$, reaching the same order of visible $VV,hh$ decay. And the other invisible decay $H_1\to N_1N_1$ maximally reaches about $0.03$ at $M_{H_1}\sim600$ GeV. The dominant production channel of $H_1$ is via gluon fusion at LHC, which can be estimated as
\begin{equation}
\sigma(gg\to H_1)\approx \sin^2\alpha\times \sigma(gg\to h),
\end{equation}
where $\sigma(gg\to h)$ is the SM Higgs production cross section but calculated with $M_h=M_{H_1}$. At present, $\sin\alpha\sim0.1$ \cite{Ilnicka:2018def} leads to the promising signatures as $H_1 \to WW\to e\nu\mu\nu$ \cite{Aaboud:2017gsl}, $ZZ\to 4\ell$ \cite{Aaboud:2018bun} and $hh\to 2b2\gamma$ \cite{Sirunyan:2018iwt,Aaboud:2018ftw}, etc. In the future, if no DM direct detection signal is observed, then the signature of heavy scalar $H_1$ will be much suppressed by tiny value of $\sin\alpha$.

\begin{table}
\begin{tabular}{|c|c|c|c|c|c|c|}
\hline
$q\bar{q}$ & $e^+e^-$  & $\mu^+\mu^-$ & $\tau^+\tau^-$& $\nu\nu$ & $NN$ & $H_1H_1$\\
\hline
~0.154~ & ~0.308~ & ~0~ & ~0.077~ & 0.192 & 0.192 & 0.077
  \\ \hline
\end{tabular}
\caption{Decay branching ratio of $U(1)_{B-2L_e-L_\tau}$ gauge boson $Z'$, where we have show the lepton flavor individually.}
\label{Tab:Zp}
\end{table}

Next, we discuss the gauge boson $Z'$ associated with $U(1)_{B-2L_e-L_\tau}$. Its decay branching ratios are flavor-dependent, which makes it quite easy to distinguish from the flavor-universal ones, such as $Z'$ from $U(1)_{B-L}$ \cite{Basso:2008iv}.
Considering the heavy $Z'$ limit, its partial decay width into fermion and scalar pairs are given by
\begin{eqnarray}
\Gamma(Z'\to f\bar{f})&=&\frac{M_{Z'}}{24\pi}g'^2 N_C^f (Q_{fL}^2+Q_{fR}^2),\\
\Gamma(Z'\to SS^*)&=&\frac{M_{Z'}}{48\pi} g'^2 Q_{S}^2,
\end{eqnarray}
where $N_C^f$ is the number of colours of the fermion $f$, i.e., $N_C^{l,\nu}=1$, $N_C^{q}=3$, and $Q_X$ is the $U(1)_{B-2L_e-L_\tau}$ charge of particle $X$. In Tab.~\ref{Tab:Zp}, we present the branching ratio of $Z'$. The dominant channel is $Z'\to e^+e^-$ with branching ratio of $0.308$, and no $Z'\to \mu^+\mu^-$. The $B-2L_e-L_\tau$ nature of $Z'$ predicts definite relation between quark and lepton final states, e.g.,
\begin{equation}
\text{BR}(Z'\to b\bar{b}):\text{BR}(Z'\to e^+e^-):\text{BR}(Z'\to \mu^+\mu^-):\text{BR}(Z'\to \tau^+\tau^-)=
\frac{1}{3}:4:0:1,
\end{equation}
which is also an intrinsic property to distinguish $Z'$ of $U(1)_{B-2L_e-L_\tau}$ from other flavored gauge bosons \cite{Chun:2018ibr}.

\begin{figure}
\begin{center}
\includegraphics[width=0.45\linewidth]{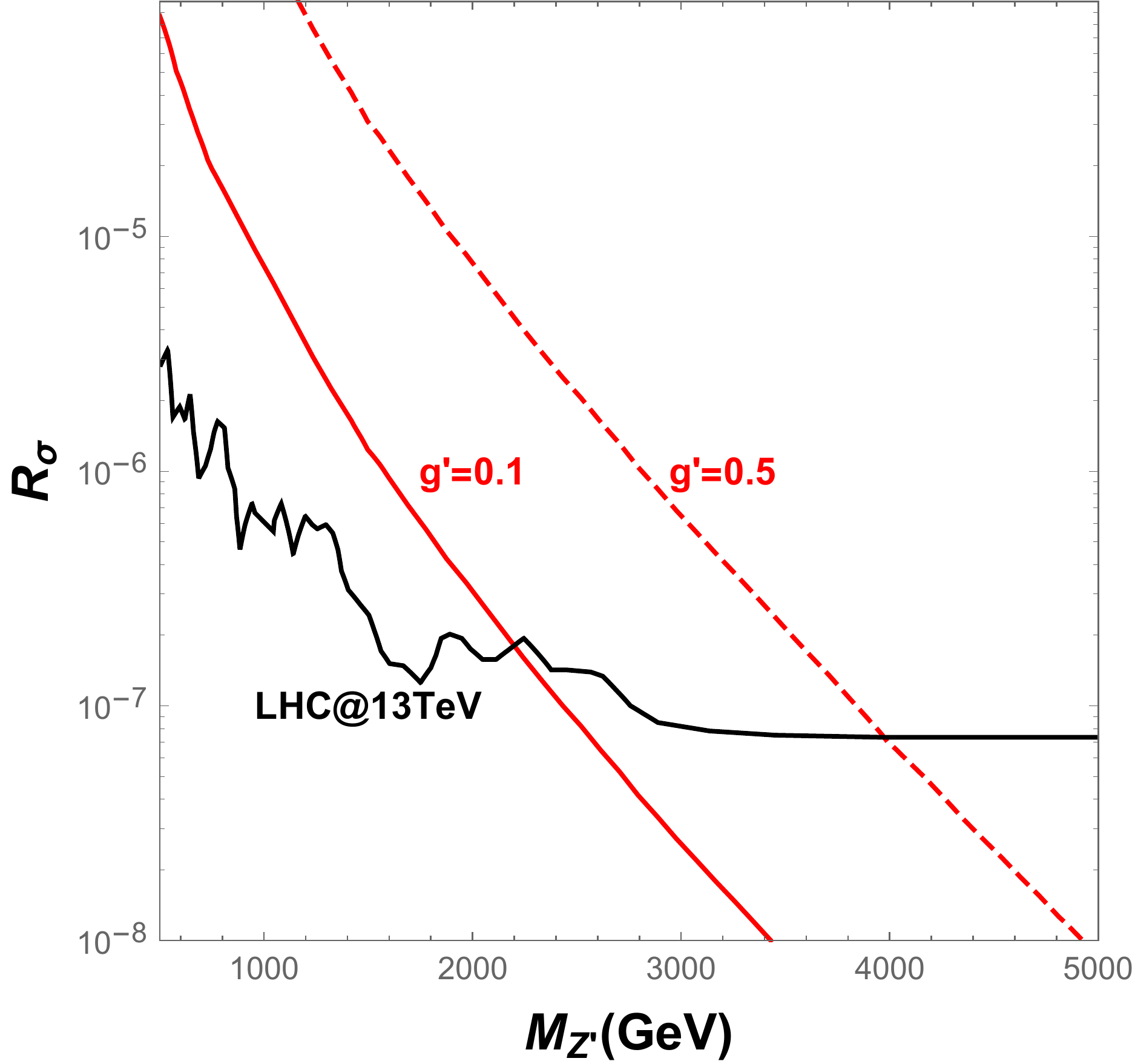}
\includegraphics[width=0.445\linewidth]{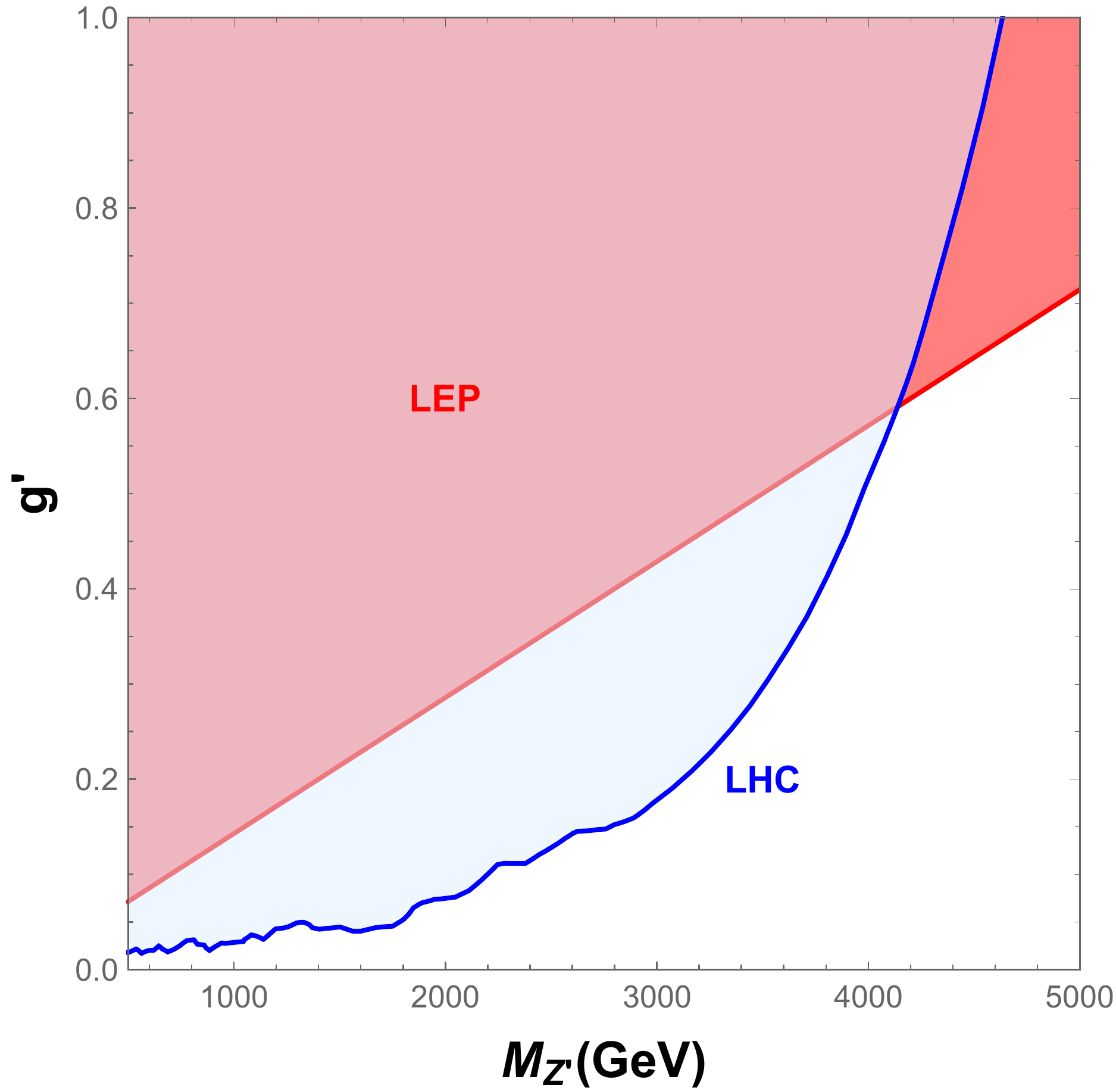}
\end{center}
\caption{ Left pattern: predicted cross section ratios in $U(1)_{B-2L_e-L_\tau}$ and corresponding limit from LHC. Right pattern: allowed parameter space in the $g'$-$M_{Z'}$ plane.
\label{Fig:LHC}}
\end{figure}

In the framework of $U(1)_{B-2L_e-L_\tau}$, one important constraint on $Z'$ comes from the precise measurement of four-fermion interactions at LEP \cite{Cacciapaglia:2006pk}, which requires
\begin{equation}\label{Eq:LEP}
\frac{M_{Z'}}{g'}\gtrsim7~\TeV.
\end{equation}
Since $Z'$ couples to both quarks and leptons, the most promising signature at LHC is the dilepton signature $pp\to Z' \to e^+e^-$. Searches for such dilepton signature have been performed by ATLAS \cite{Aaboud:2017buh} and CMS collaboration \cite{Sirunyan:2018exx}. Because of no $Z'\to \mu^+\mu^-$ channel, we can only take the results from CMS, which provides a limit on the ratio
\begin{equation}
R_\sigma=\frac{\sigma(pp\to Z'+X\to e^+e^-+X)}{\sigma(pp\to Z+X\to e^+e^-+X)}.
\end{equation}
The theoretical cross section of the dilepton signature are calculated by using {\tt MadGraph5\_aMC@NLO} \cite{Alwall:2014hca}. Left pattern of Fig.~\ref{Fig:LHC} shows that the dilepton signature has excluded $M_{Z'}\lesssim2.2(4.0)$ TeV for $g'=0.1(0.5)$. Then comparing the theoretical ratio with experimental limit, one can acquire the exclusion limit in the $g'-M_{Z'}$ plane as shown in right pattern of Fig.~\ref{Fig:LHC}. Obviously, LHC limit is more stringent than LEP when $M_{Z'}\lesssim4$ TeV.

The inert charge scalars $\eta_{1,2}^\pm$ are also observable at LHC. They can decay into charged leptons and right-hand singlets via the Yukawa interactions as
\begin{eqnarray}
\Gamma(\eta^\pm_1\to \ell^\pm N_i)&=&\frac{M_{\eta_1^\pm}}{16\pi} \left|h'_{\ell i}\right|^2\left(1-\frac{M_{N_i}^2}{M_{\eta^\pm_1}^2}\right)^2,\\
\Gamma(\eta^\pm_2\to \ell^\pm N_i)&=&\frac{M_{\eta_1^\pm}}{16\pi} \left|f'_{\ell i}\right|^2\left(1-\frac{M_{N_i}^2}{M_{\eta^\pm_2}^2}\right)^2.
\end{eqnarray}
From Eq. \eqref{Eq:Yuk}, we aware that $\eta_1^\pm$ decays into $\mu,\tau$ final states, while $\eta_2^\pm$ decays into $e,\tau$ final states. The electron-phobic nature of $\eta_1^\pm$ and muon-phobic nature of $\eta_2^\pm$ make them quite easy to distinguish. Meanwhile, their decay branching ratios are related by neutrino oscillation data through the Yukawa coupling $h',f'$. Considering the benchmark point in Eq. \eqref{Eq:BF}, the predicted branching ratios are shown in Tab. ~\ref{TB:eta} in the heavy scalar limit. The dominant decay channel of $\eta^\pm_1$ is $\mu^\pm N_1$, and $\tau^\pm N_1$ for $\eta^\pm_2$. So $\eta^\pm_1$ is expected easier to be discovered.
Produced via the Drell-Yan process $pp\to \eta^+_1\eta^-_1,\eta^+_2\eta^-_2$, the decay channel $\eta^\pm_{1,2} \to \ell^\pm N_1$ then leads to signature $\ell^+\ell^-+\cancel{E}_T$. Exclusion region by direct LHC search  for such signature \cite{Aaboud:2018jiw} has been shown in right pattern of Fig.~\ref{Fig:CB}. To satisfy the direct LHC search bounds, one needs either $M_{N_1}\lesssim M_{\eta}<500$ GeV or $M_{\eta}>500$ GeV.

\begin{center}
\begin{table}
\begin{tabular}{|c||c|c|c|c|c|c|}
\hline
Final state & $e^\pm N_1$ & $\mu^\pm N_1$ & $\tau^\pm N_1$
          & $e^\pm N_2$ & $\mu^\pm N_2$ & $\tau^\pm N_2$
\\ \hline
$\eta^\pm_1$ & 0.000 & 0.465 & 0.099 & 0.000 & 0.178 & 0.258
\\ \hline
$\eta^\pm_2$ & 0.043 & 0.000 & 0.611 & 0.113 & 0.000 & 0.233
\\ \hline
\end{tabular}
\caption{Branching ratios of charge scalar $\eta_{1,2}^\pm$.}
\label{TB:eta}
\end{table}
\end{center}

\section{Conclusion}\label{Sec:CL}

The scotogenic model is an elegant pathway to explain the origin of neutrino mass and dark matter. Meanwhile, texture-zeros in neutrino mass matrix provide a promising way to under stand the leptonic flavor structure. Therefore, it is appealing to connect the scotogenic model with texture-zeros. In this paper, we propose a viable approach to realise two texture-zeros in the scotogenic model with flavor dependent $U(1)_{B-2L_\alpha-L_\beta}$ gauge symmetry. These models are extended by two right-handed singlets $N_{Ri}$ and two inert scalar doublets $\eta_{i}$, which are odd under the dark $Z_2$ symmetry. Six kinds of texture-zeros are realised in our approach, i.e., texture $A_1$, $A_2$, $B_3$, $B_4$, $D_1$ and $D_2$. Among all the six texture-zeros, we find that texture $A_1$ and $A_2$ are allowed by current experimental limits, while texture $B_3$ and $B_4$ are marginally allowed. Besides, texture $D_1$ and $D_2$ are already excluded by neutrino oscillation data.

Realization of texture-zeros in the scotogenic model makes the model quite predictive. And we have taken texture $A_1$ derived from $U(1)_{B-2L_e-L_\tau}$ for illustration. Some distinct features are summarized in the following:
\begin{itemize}
  \item The texture $A_1$ predicts vanishing neutrinoless double beta decay rate. And only normal neutrino mass hierarchy is allowed.
      It predicts $m_1\sim 0.007$ eV, $m_2\sim 0.01$ eV, and $m_3\approx\sqrt{\Delta m^2}\sim 0.05$ eV, then $\sum m_i\sim 0.07$ eV. There are also strong correlation between the Dirac and Majorana phases, i.e., $\rho\approx\frac{\delta}{2}$ and $\sigma\approx\frac{\delta}{2}-\frac{\pi}{2}$.
  \item The ratios of corresponding  Yukawa couplings are also predicted by neutrino oscillation data, e.g.,
      \begin{eqnarray}\nonumber
      \frac{h_{\tau2}}{h_{\mu1}}:\frac{f_{\tau1}}{h_{\mu1}}
      :\frac{f_{e2}}{h_{\mu1}}\simeq 0.745:0.933:0.401.
      \end{eqnarray}
  \item Due to specific Yukawa structure, the LFV process $\mu\to e\gamma$ is missing at one-loop level. Meanwhile, large cancellations are possible for $\tau\to \mu\gamma$ and $\tau \to e\gamma$ with degenerate right-handed singlets. More stringent constraint comes from muon anomalous magnetic moment $\Delta a_\mu$. Although $\mathcal{O}(1)$ Yukawa couplings are easily to avoid such limit.
  \item Satisfying all constraints, correct relic density of dark matter $N_1$ is achieved for $M_{N_1}\lesssim M_{\eta}<500$ GeV with $h_{\mu1}\lesssim 1$ or $M_\eta>500$ GeV with $h_{\mu1}>1$.As for direct detection, we have shown that the predicted spin-independent DM-nucleon cross section $\sigma^\text{SI}$ with $\sin\alpha=0.1$ satisfies the current XENON1T limit, but is within future reach of LZ.
  \item The massless Mojoron $J$ contributes to invisible decay of SM Higgs. The additional scalr singlet $H_1$ can be probe in the channel $gg\to H_1\to W^+W^-, ZZ$ at LHC. Decays of charged scalars $\eta_{1,2}^\pm$ lead to $pp\to \eta^+_{1,2}\eta^-_{1,2}\to \ell^+\ell^-+\cancel{E}_T$ signature. Note that the corresponding branching ratios are also correlated with neutrino oscillation parameters.
  \item The neutral gauge boson $Z'$ is promising via the di-electron signature $pp\to Z' \to e^+e^-$. Its $B-2L_e-L_\tau$ nature can be confirmed by
      \begin{equation}\nonumber
      \text{BR}(Z'\to b\bar{b}):\text{BR}(Z'\to e^+e^-):\text{BR}(Z'\to \mu^+\mu^-):\text{BR}(Z'\to \tau^+\tau^-)=\frac{1}{3}:4:0:1,
      \end{equation} 
\end{itemize}

In a nutshell, the scotogenic model with flavor dependent $U(1)_{B-2L_\alpha-L_\beta}$ symmetry predicts distinct and observable phenomenology, which is useful to distinguish from other models.
\section{Acknowledgements}

The work of Weijian Wang is supported by National Natural Science Foundation of China under Grant Numbers 11505062, Special Fund of Theoretical Physics under Grant Numbers 11447117 and Fundamental Research Funds for the Central Universities under Grant Numbers 2014ZD42. The work of Zhi-Long Han is supported by National Natural Science Foundation of China under Grant No. 11805081 and No. 11605075, Natural Science Foundation of Shandong Province under Grant No. ZR2018MA047, No. ZR2017JL006 and No. ZR2014AM016.


\end{document}